\documentclass[twocolumn]{aastex631}

\usepackage{natbib}
\usepackage{amsmath}
\usepackage{url}
\usepackage{graphicx}

\begin{document}

\title{Imaging Polarization of the Blue-Excess Hot Dust-Obscured Galaxy WISE J011601.41--050504.0}

\author[0000-0002-9508-3667]{R.J.~Assef}
\affiliation{N\'ucleo de Astronom\'ia de la Facultad de Ingenier\'ia y Ciencias, Universidad Diego Portales, Av. Ej\'ercito Libertador 441, Santiago, Chile}

\author[0000-0002-8686-8737]{F.E.~Bauer}
\affiliation{Instituto de Astrof\'isica, Facultad de F\'isica, Pontificia Universidad Cat\'olica de Chile, 306, Santiago 22, Chile}
\affiliation{Millennium Institute of Astrophysics (MAS), Nuncio Monse\~nor S\'otero Sanz 100, Providencia, Santiago, Chile}
\affiliation{Space Science Institute, 4750 Walnut Street, Suite 205, Boulder, Colorado 80301, USA}

\author[0000-0001-7489-5167]{A.W.~Blain}
\affiliation{Physics \& Astronomy, University of Leicester, 1 University Road, Leicester LE1 7RH, UK}

\author[0000-0002-8147-2602]{M.~Brightman}
\affiliation{Cahill Center for Astrophysics, California Institute of Technology, 1216 East California Boulevard, Pasadena, CA 91125, USA}

\author[0000-0003-0699-6083]{T.~D\'iaz-Santos}
\affiliation{Institute of Astrophysics, Foundation for Research and Technology–Hellas (FORTH), Heraklion, GR-70013, Greece}

\author{P.R.M.~Eisenhardt}
\affiliation{Jet Propulsion Laboratory, California Institute of Technology, 4800 Oak Grove Drive, Pasadena, CA 91109, USA}
  
\author[0000-0003-1470-5901]{H.D.~Jun}
\affiliation{SNU Astronomy Research Center, Astronomy Program, Dept. of Physics and Astronomy, Seoul National University, Seoul 08826, Republic of Korea}
  
\author[0000-0003-2686-9241]{D.~Stern}
\affiliation{Jet Propulsion Laboratory, California Institute of Technology, 4800 Oak Grove Drive, Pasadena, CA 91109, USA}

\author[0000-0002-9390-9672]{C.-W.~Tsai}
\affiliation{National Astronomical Observatories, Chinese Academy of Sciences, 20A Datun Road, Chaoyang District, Beijing, 100012, People's Republic of China}

\author[0000-0001-5819-3552]{D.J.~Walton}
\affiliation{Centre for Astrophysics Research, University of Hertfordshire, College Lane, Hatfield AL10 9AB, UK}

\author[0000-0001-7808-3756]{J.W.~Wu}
\affiliation{University of Chinese Academy of Sciences, Beijing 100049, People's Republic of China}
\affiliation{National Astronomical Observatories, Chinese Academy of Sciences, 20A Datun Road, Chaoyang District, Beijing, 100012, People's Republic of China}

\begin{abstract}
We report on VLT/FORS2 imaging polarimetry observations in the $R_{\rm special}$ band of WISE J011601.41--050504.0 (W0116--0505), a heavily obscured hyper-luminous quasar at $z=3.173$ classified as a Hot, Dust-Obscured Galaxy (Hot DOG) based on its mid-IR colors. Recently, \citet{assef20} identified W0116--0505 as having excess rest-frame optical/UV emission, and concluded this excess emission is most likely scattered light from the heavily obscured AGN. We find that the broad-band rest-frame UV flux is strongly linearly polarized (10.8$\pm$1.9\%, with a polarization angle of 74$\pm$9~deg), confirming this conclusion. We analyze these observations in the context of a simple model based on scattering either by free electrons or by optically thin dust, assuming a classical dust torus with polar openings. Both can replicate the degree of polarization and the luminosity of the scattered component for a range of geometries and column densities, but we argue that optically thin dust in the ISM is the more likely scenario. We also explore the possibility that the scattering medium corresponds to an outflow recently identified for W0116--0505. This is a feasible option if the outflow component is bi-conical with most of the scattering occurring at the base of the receding outflow. In this scenario the quasar would still be obscured even if viewed face on, but might appear as a reddened type 1 quasar once the outflow has expanded. We discuss a possible connection between blue-excess Hot DOGs, extremely red quasars (ERQs), reddened type 1 quasars, and unreddened quasars that depends on a combination of evolution and viewing geometry.
\end{abstract}


\section{Introduction}\label{sec:intro}

Hot Dust-Obscured Galaxies (Hot DOGs) are a population of hyper-luminous obscured quasars \citep{eisenhardt12, wu12} identified by NASA's {\it{Wide-field Infrared Explorer}} \citep[WISE;][]{wright10}. Hot DOGs comprise some of the most luminous galaxies in the Universe, most with $L_{\rm Bol}\gtrsim 10^{13}~L_{\odot}$ and $\sim 10\%$ with bolometric luminosities exceeding $10^{14}~L_{\odot}$, without signs of gravitational lensing \citep{tsai15}. A number of studies have identified a hyper-luminous, highly-obscured AGN as the primary source of the luminosity in these objects \citep[e.g.,][]{eisenhardt12, assef15}. This obscured AGN component dominates at mid-IR wavelengths \citep{assef15}, but is sometimes luminous enough to dominate the emission at far-IR wavelengths as well \citep{jones14, diaz-santos16, diaz-santos21}. Hard X-ray spectra have been obtained for several Hot DOGs, leading to the conclusion that the obscuration is close to, or above, the Compton-thick threshold \citep{stern14, piconcelli15, assef16, assef20, vito18}. Recent studies have shown that some Hot DOGs are driving massive outflows of ionized gas \citep{diaz-santos16, jun20, finnerty20} as well as possibly molecular gas outflows \citep{fan18}, suggesting the obscured, hyper-luminous quasar may be in the course of shutting down star-formation by removing the host-galaxy gas reservoir. Additionally, a number of studies have identified that at least part of the Hot DOG population may be involved in mergers \citep{fan16, farrah17, assef20}, with the clearest case being the discovery that the most luminous Hot DOG may be at the center of a multiple merger system with three neighboring galaxies \citep{diaz-santos18}. This could make Hot DOGs consistent with being at the blow-out stage of the massive galaxy evolution scheme suggested by, e.g., \citet{hopkins08}, although \citet{diaz-santos21} has recently suggested that the Hot DOG phase may be recurrent throughout the lifetime of a massive galaxy.

Hot DOGs have very distinctive UV through IR SEDs \citep[e.g., see][]{tsai15, assef15}. The highly-obscured, hyper-luminous AGN dominates the IR SEDs of these objects as well as the bolometric luminosity output, while a moderately star-forming galaxy without significant obscuration typically dominates the UV and optical portions of the SED \citep[see, e.g.,][]{eisenhardt12, jones14, assef15}. \citet{assef16} studied the UV through mid-IR SED of a large number of Hot DOGs and identified 8 objects that showed considerably bluer SEDs. They dubbed this sub-sample blue-excess Hot DOGs, or BHDs. They found that the excess blue emission had a power-law shape and was best modeled by the emission from an unobscured or lightly-obscured AGN accretion disk with $\sim$1\% of the luminosity of the obscured AGN responsible for the IR emission. \citet{assef16} discussed the possible origins of this excess blue emission, and presented a detailed study of one BHD, WISE J020446.13--050640.8 (W0204--0506 hereafter, $z=2.100$). They concluded that the most likely source of the blue-excess emission in W0204--0506 is light from the highly-obscured, hyper-luminous AGN scattered into our line of sight. In a follow-up study, \citet{assef20} presented further observations of this source, as well as a detailed study of two more BHDs, WISE J011601.41--050504.0 (W0116--0505 hereafter, $z=3.173$) and WISE J022052.12+013711.6 (W0220+0137 hereafter, $z=3.122$). They also concluded that the most likely source of the excess blue emission in all three sources is scattered light from the obscured AGN, although a contribution from star-formation could not be completely ruled out, particularly in the case of W0204--0506. 

A possible way to differentiate between the star-formation and scattered-light origins for the blue excess is through linear polarization in the UV, since a high degree of polarization would be expected for the latter scenario, but not for unobscured star-formation. There is a rich history of detecting scattered emission in obscured AGN through their linearly polarized flux, and such observations comprise the basis of the AGN unification model \citep[see, e.g.,][]{antonucci85, miller91, antonucci93}. Furthermore, unobscured quasars do not show high polarization. \citet{berriman90} measured the polarization properties of 114 type 1 QSOs in the Palomar-Green Quasar Survey \citep{schmidt83} and found an average polarization of 0.5\% with a maximum of 2.5\%. Luminous obscured quasars and radio galaxies, on the other hand, show significantly larger polarization, up to $\sim20\%$ in some cases \citep[see, e.g.,][]{hines95, smith00, vernet01, zakamska05, alexandroff18}.

In this article we present imaging polarimetric observations of W0116--0505, the brightest of the BHDs studied by \cite{assef20} at observed-frame optical wavelengths, carried out with the FOcal Reducer/low dispersion Spectrograph 2 (FORS2) instrument at the Very Large Telescope (VLT), finding high linear polarization and confirming that scattered light emission is the source of the blue-excess in this BHD. In \S\ref{sec:observations} we describe these observations as well as other supporting observations presented by previous studies. In \S\ref{sec:analysis} we discuss our linear polarization measurements. In \S\ref{sec:discussion} we discuss in more detail the supporting indirect evidence for scattered light in BHDs, and discuss the implications of the linear polarization detection on the obscuration geometry and scattering medium. Finally, in \S\ref{sec:conclusions} we summarize our conclusions. All errors are quoted at the 1-$\sigma$ level. Throughout the article we assume a flat $\Lambda$CDM cosmology with $H_0=67.7~\rm km~\rm s^{-1}~\rm Mpc^{-1}$ and $\Omega_M=0.307$ \citep{planck16}.

\section{Observations}\label{sec:observations}

\subsection{Photometric and Spectrocopic Data}\label{ssec:prev_data}

\begin{figure*}
  \plotone{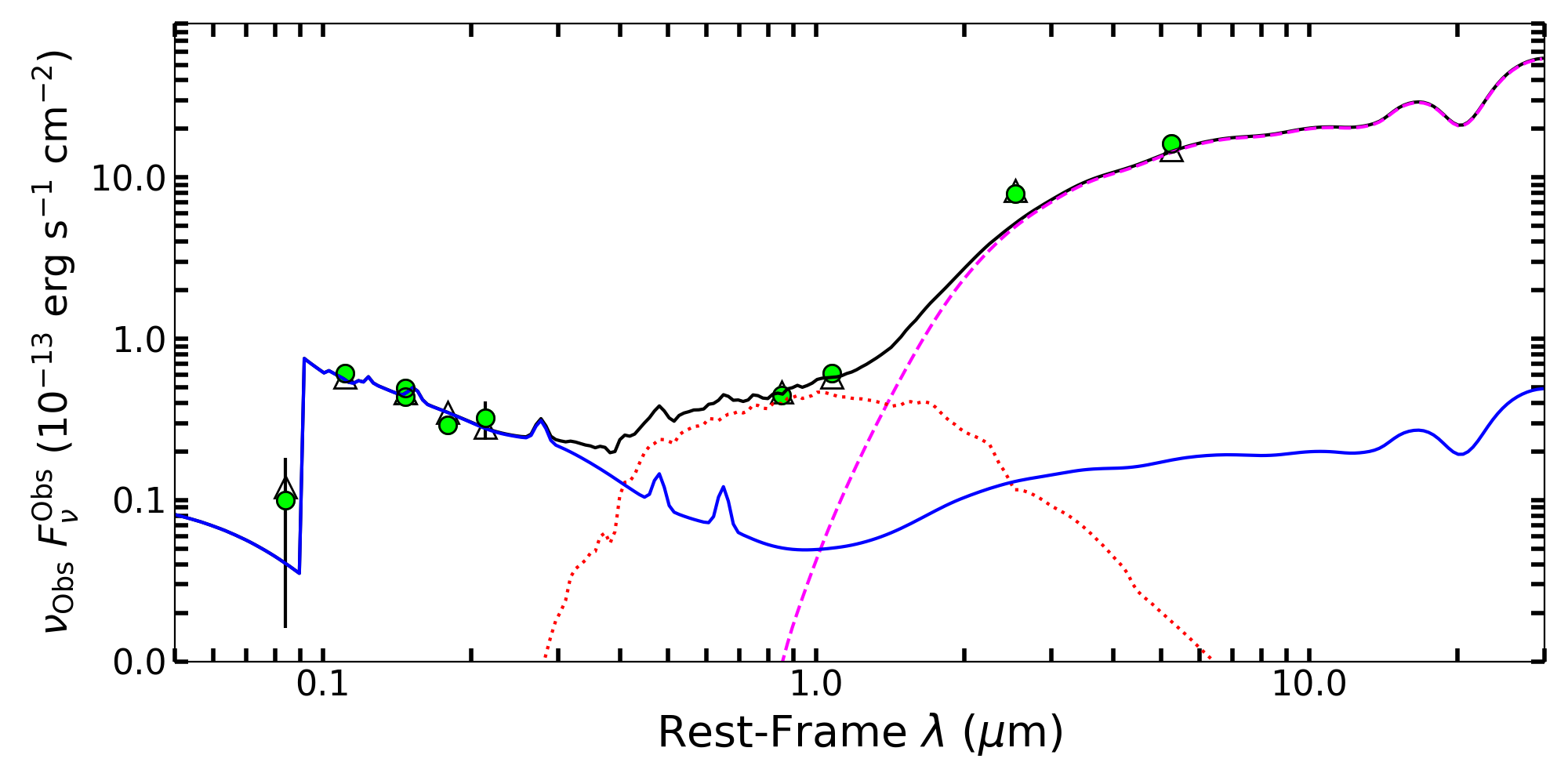}
  \caption{SED of W0116--0505. The green circles show the observed fluxes in the SDSS $u^{\prime}g^{\prime}r^{\prime}i^{\prime}z^{\prime}$ bands, the WISE W1--4 bands and the SOAR $r$-band presented by \citet{assef15}. The solid black line shows the best-fit SED model, with the primary, obscured AGN SED component shown by the dashed magenta line, and the secondary, unobscured AGN SED component shown by the blue solid line. The dotted red line shows the host galaxy component. The secondary AGN component most likely corresponds to scattered emission from the primary AGN component (see text for details).}
  \label{fg:sed}
\end{figure*}

W0116--0505 was identified by \citet{assef20} as a BHD from its multi-wavelength SED, shown in Figure \ref{fg:sed}. The figure also shows the best-fit model SED obtained using the templates and algorithm of \citet{assef10} but modified to consider a second AGN SED component (see \citealt{assef16, assef20} for details). Briefly, the SED is modeled as a non-negative linear combination of three empirically determined host galaxy templates (referred to as E, Sbc and Im as they correspond to modified versions of the \citealt{cww80} templates of the same name) and two AGN components. Each AGN SED component uses the same underlying template but with independent luminosity and obscuration. The latter is quantified by the color excess $E(B-V)$ and assumes $R_V=3.1$ and a reddening law equal to that of the SMC of \citet{gordon98} at $\lambda<3300$\AA\ and that of the Milky-Way of \citet{cardelli89} at longer wavelengths (see \citealt{assef10} for further details). The obscured, more luminous AGN component has intrinsic $\log L_{6\mu\rm m}/(\rm erg~\rm s^{-1}) = 47.24^{+0.17}_{-0.11}$ and $E(B-V)=4.24^{+2.71}_{-1.23}~\rm mag$, while the lower luminosity, unobscured AGN component has intrinsic $\log L_{6\mu\rm m}/(\rm erg~\rm s^{-1}) = 45.18^{+0.04}_{-0.03}$ and $E(B-V)<0.02~\rm mag$. 

\begin{figure*}
  \plotone{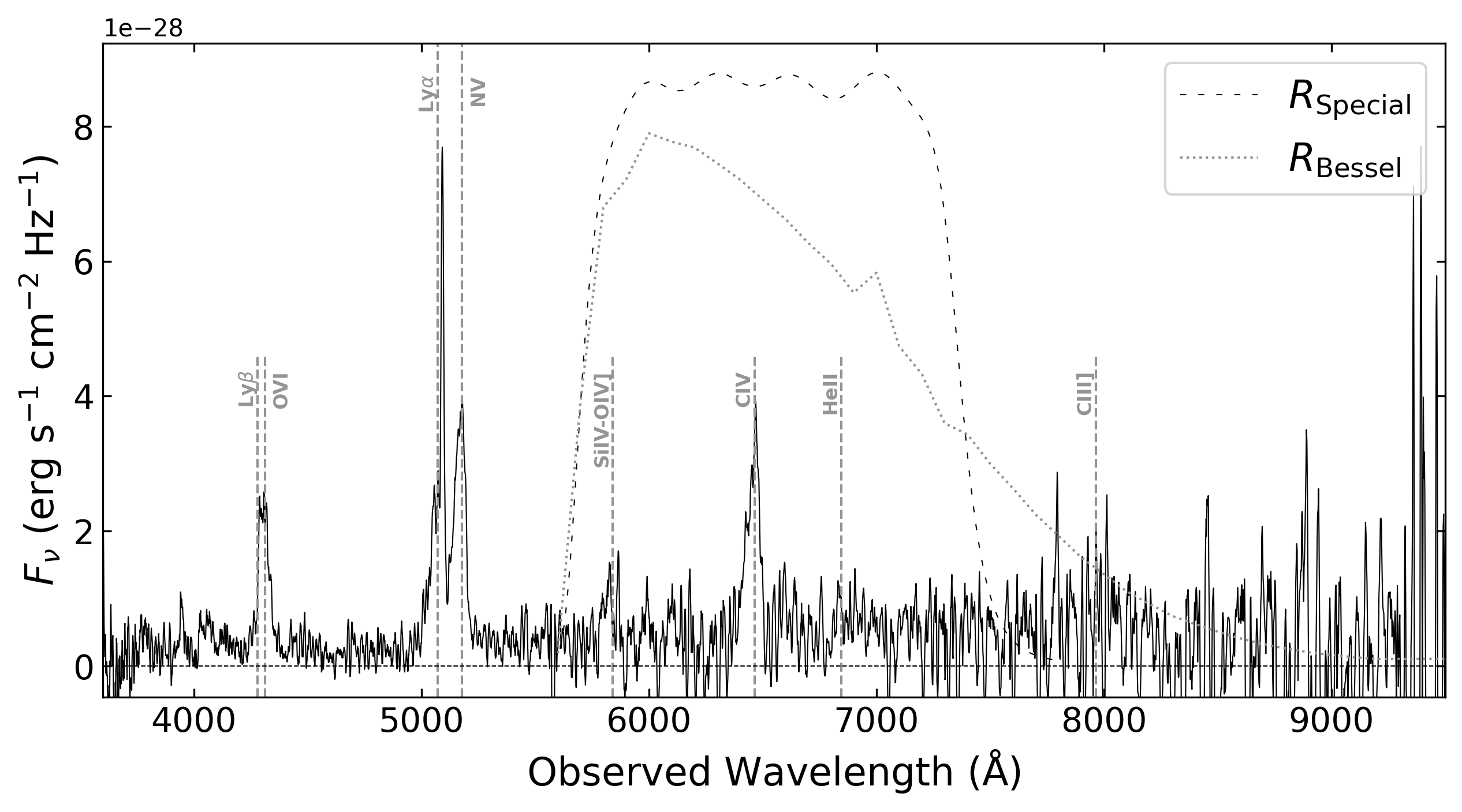}
  \caption{SDSS spectrum of W0116--0505. The emission lines are marked with vertical dashed gray lines assuming a redshift of 3.173  (see \S\ref{ssec:prev_data} for details about the redshift). The plot also shows the response function of the FORS $R_{\rm Special}$ (black dashed line) and $R_{\rm Bessel}$ (gray dotted line) filters.}
  \label{fg:spec_and_filters}
\end{figure*}

W0116--0505 was observed spectroscopically by SDSS \citep{eisenstein11}. The spectrum, shown in part in Figure \ref{fg:spec_and_filters} \citep[see Fig.2 of][for the full spectrum]{assef20}, displays emission lines typically associated with quasars, supporting the scenario in which the blue-excess emission is due to scattered light from the obscured hyper-luminous quasar \citep[see][for details]{assef20}. A closer look into some of the emission lines further reinforces this case. Specifically, by modeling the C{\sc iv} emission line and the Si{\sc iv}-O{\sc iv}] blend with single Gaussian functions and a linear local continuum, as described in \citet{assef20}, we find that their flux ratio of $2.9\pm 0.8$ is consistent with the ratio of 2.8 found in the \citet{vandenberk01} quasar composite. The equivalent width (EW) of the Si{\sc iv}-O{\sc iv}] of 32$\pm$23\AA\ is not well constrained, yet consistent with the value of 8.13\AA\ in the quasar composite. For C{\sc iv}, on the other hand, we find a larger EW of 86$\pm$17\AA\ compared with the 23.78\AA\ in the quasar composite, but well in line with the mean of the distribution of C{\sc iv} EWs found by \citet[][see their Fig. 13]{rakshit20} for the SDSS DR14 quasar sample. We also model the Ly$\beta$-O{\sc vi} blend with two Gaussian functions, as well as the Ly$\alpha$-N{\sc v} with three Gaussian functions (as an extra narrow component seems to be needed for Ly$\alpha$). We find that the combined flux ratio of Ly$\alpha$-N{\sc v} with respect to C{\sc iv}  of 4.38$\pm$0.37 is consistent with that of 4.1 found in the \citet{vandenberk01} composite. We find, however, that the Ly$\beta$-O{\sc vi} blend has a combined flux with respect to C{\sc iv} of 1.56$\pm$0.14, significantly in excess of the ratio of 0.38 found in the quasar composite. This may indicate specific properties of the IGM/CGM around W0116--0505. A detailed characterization of emission lines in Hot DOGs is being prepared by \citet{eisenhardt22}, and should help elucidate whether this is a common feature among these objects. 

\begin{figure*}
  \plotone{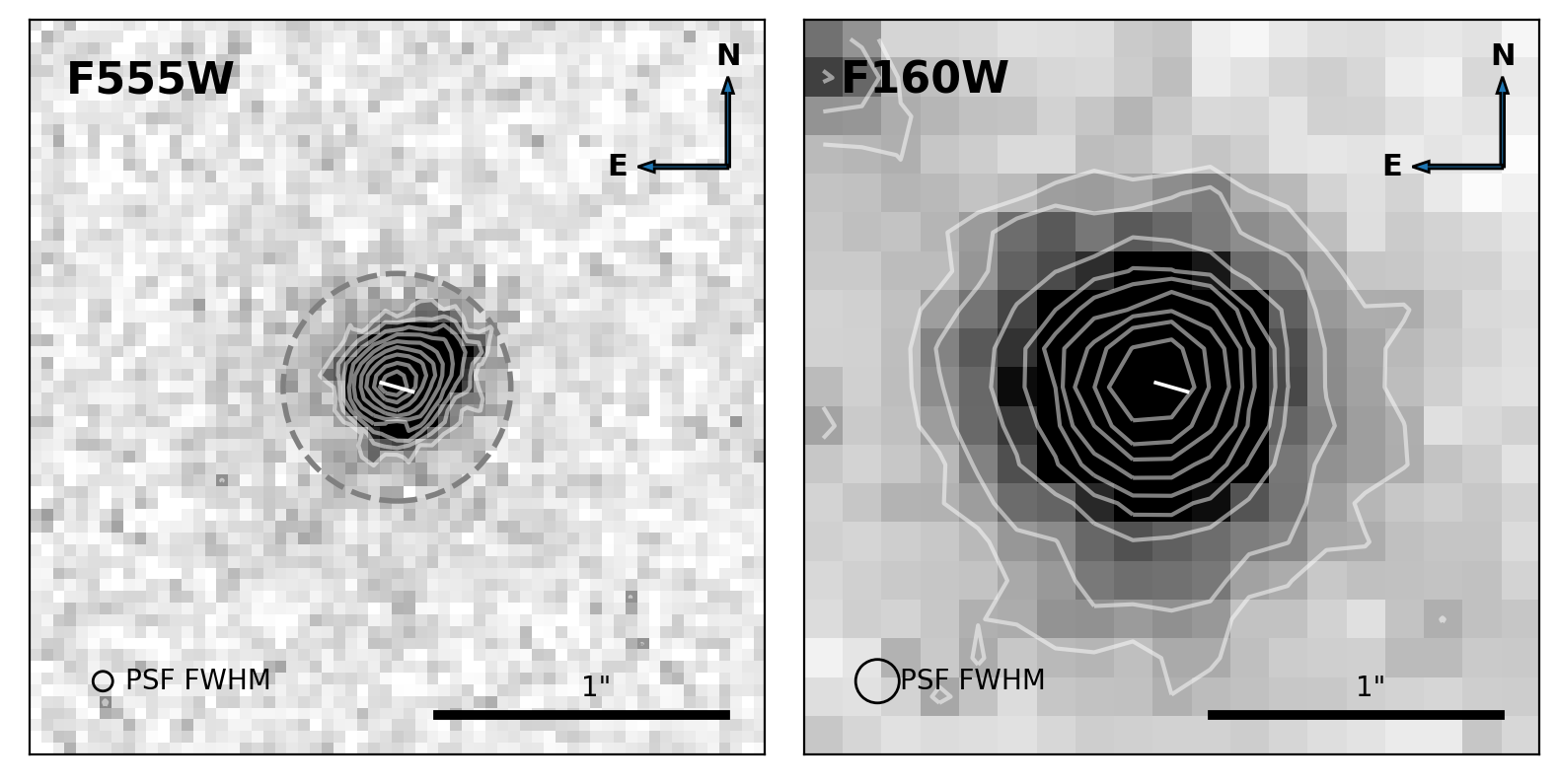}
  \caption{Cutouts (2.5\arcsec$\times$2.5\arcsec) centered on W0116--0505 in the {\it{HST}}/WFC3 F555W (left) and F160W (right) bands. The white line in the center shows the orientation of the linear polarization measured in \S\ref{sec:analysis}. The white contours show logarithmically spaced levels in relation to the brightest pixel, starting at 1\%. The circles at the bottom left of each panel show the size of the PSF FWHM for each band. The gray dashed circle in the left panel is centered on W0116--0505 and has a physical radius of 3~kpc, as discussed in \S\ref{ssec:outflow}.}
  \label{fg:HST}
\end{figure*}

SDSS reported the redshift to be 3.1818$\pm$0.0006. \citet{wu12}, using a spectrum obtained at the MMT observatory, reported a slightly lower redshift of 3.173$\pm$0.001. A close inspection of the data shows that while the peak of the narrow component of the Ly$\alpha$ emission is consistent with the SDSS redshift, the O{\sc vi} and the broad component of the Ly$\alpha$ emission line are consistent with that of \citet{wu12}. Recently, \citet{diaz-santos21} used a spectrum covering the CO(4--3) emission line obtained with ALMA to constrain the systemic redshift of W0116--0505 to 3.1904$\pm$0.0002, implying a significant blue shift of the UV emission lines of $615\pm 45$ and $1245\pm 73~\rm km~\rm s^{-1}$, respectively, for the redshift estimates from SDSS and \citet{wu12}. While a significant blue shift could be indicative of an ongoing merger, \citet{assef20} determined the morphology of the W0116--0505 host galaxy in the {\it{HST}}/WFC3 F160W band is consistent with an undisturbed early-type galaxy. This image, along with the {\it{HST}}/WFC3 F555W image presented by \citet{assef20}, is shown in Figure \ref{fg:HST}. Note that, as discussed by \citet{assef20}, the emission is resolved in both HST bands. We find that the source has half-light radii of 0.11\arcsec\ (0.9~kpc) and 0.23\arcsec\ (1.8~kpc) in the F555W and F160W bands respectively. For reference, the PSF in those bands respectively have FWHM of .067\arcsec\ and 0.148\arcsec.

\subsection{Imaging Polarimetry}\label{ssec:impol}

\begin{figure*}
  \plotone{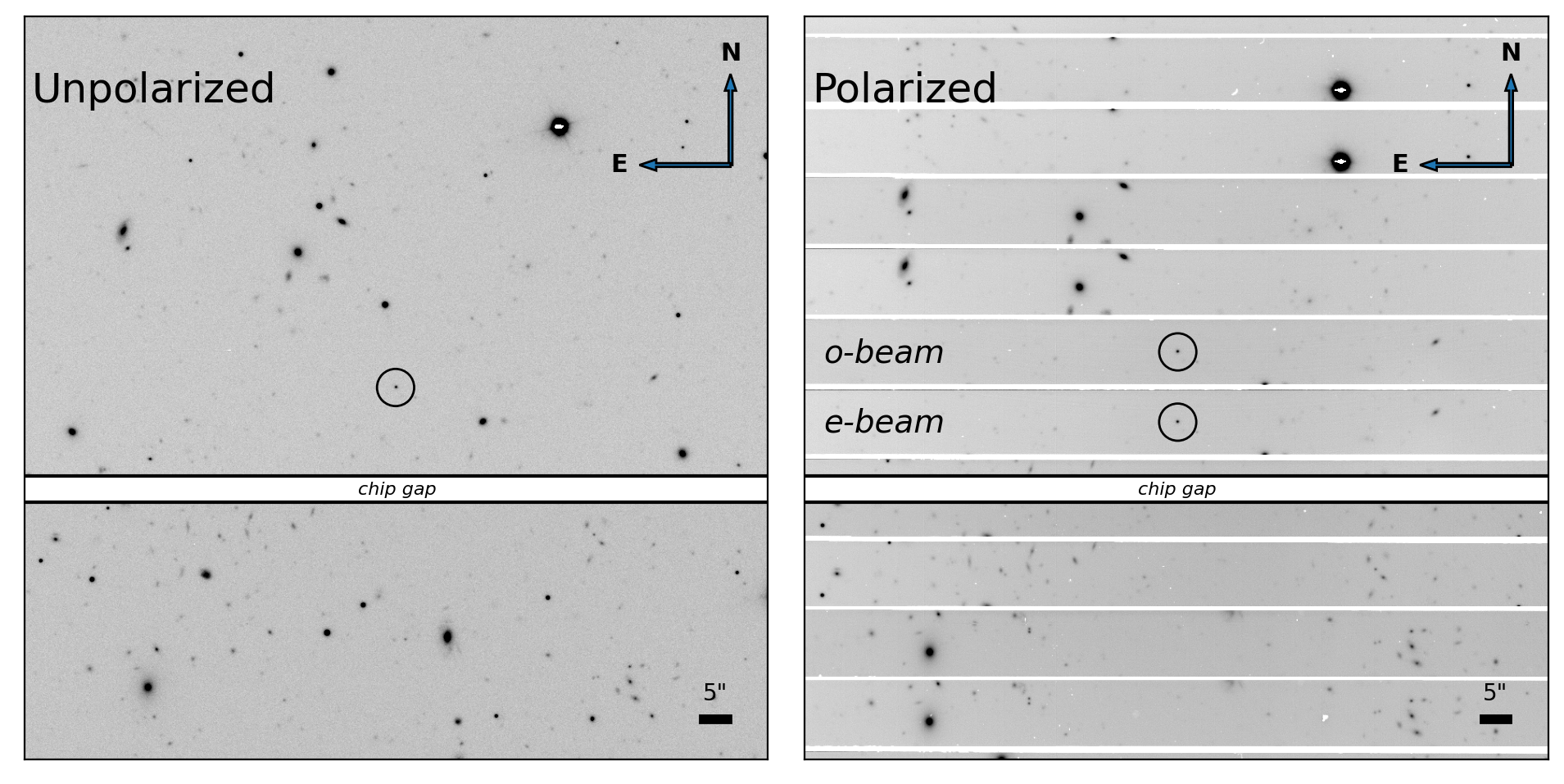}
  \caption{Cutouts (2\arcmin$\times$2\arcmin) of the VLT/FORS2 images of W0116--0505 obtained on MJD 59135. The left panel shows an image in the $R_{\rm Special}$ band without the polarization optics, while the right panel shows a stack of all the images obtained that night using the polarization optics. Note that a mask is used to separate the $e$- and $o$-beams, which alternate from top to bottom, allowing only 50\% of the FoV to be observed at a time. The 3\arcsec\ radius black circle shows the position of W0116--0505.}
  \label{fg:pol_unpol_images}
\end{figure*}

Imaging polarimetric observations of W0116--0505 were obtained with the FORS2 instrument at the VLT using the $R_{\rm Special}$ broad-band filter ($5710-7360$~\AA, see Fig. \ref{fg:spec_and_filters}). The observations were divided into four equal observing blocks (OBs), one executed on UT~2020-10-13, two on UT~2020-10-14, and the remaining one on UT~2020-10-15. Each observing block consisted of 2$\times$353s observations of the target at retarder plate angles of 0, 22.5, 45 and 67.5 degrees, which are the recommended angles for linear polarization measurements in the FORS2 User Manual\footnote{\url{https://www.eso.org/sci/facilities/paranal/instruments/fors/doc/VLT-MAN-ESO-13100-1543\_P06.pdf}}. A zero polarization standard star, WD 2039--202, and a polarization standard star, BD-12 5133, where observed with the same configuration on the night of UT 2020-10-12. Another zero polarization standard, WD 0310--688, was observed with the same configuration on the night of UT 2020-09-25. Figure \ref{fg:pol_unpol_images} shows an example of the images obtained for W0116--0505 on the first night of observations. Note that FORS2 simultaneously observes the ordinary ($o$) and extra-ordinary ($e$) beams for point sources \citep[see][as well as the FORS2 User Manual for further details]{gonzalez20}. Because of this, the standard stars observed are not needed for calibration purposes to measure linear polarization, but are still useful to check for the presence of systematic offsets. Table \ref{tab:obslog} summarizes the observations.

\begin{deluxetable*}{lccccc}

    \tablecaption{Summary of VLT/FORS2 Imaging Polarimetry Observations\label{tab:obslog}}
    
    \tablehead{
        \colhead{Target} &
        \colhead{Program ID} &
        \colhead{OB} & 
        \colhead{Mean MJD} &
        \colhead{Airmass} &
        \colhead{Seeing}\\
        \colhead{} &
        \colhead{} &
        \colhead{} &
        \colhead{(days)} &
        \colhead{} &
        \colhead{(\arcsec)}
    }
    
    \tabletypesize{\small}
    \tablewidth{0pt}
    \tablecolumns{6}
    
    \startdata
    W0116-0505  &  106.218J.001  & \phn\phn2886768 & 59135.18 & 1.064 & 0.57\\
                &                & \phn\phn2886765 & 59136.17 & 1.069 & 0.84\\
                &                & \phn\phn2886772 & 59136.20 & 1.073 & 0.85\\
                &                & \phn\phn2886622 & 59137.17 & 1.066 & 0.91\\
    \\
    \multicolumn{6}{l}{{\it{Polarization Standard}}}\\
    BD-12 5133  &  60.A-9203(E)  &       200277970 & 59133.98 & 1.117 & 1.19\\
    \\
    \multicolumn{6}{l}{{\it{Zero-polarization Standards}}}\\    
    WD 0310-688 &  60.A-9203(E)  &       200277988 & 59117.36 & 1.410 & 1.58\\
    WD 2039-202 &  60.A-9203(E)  &       200278006 & 59134.00 & 1.003 & 0.70\\
    \enddata
    
\end{deluxetable*}

All images were bias subtracted using the EsoRex pipeline\footnote{\url{https://www.eso.org/sci/software/cpl/esorex.html}}. We did not apply a flat-field correction as it is not recommended for linear polarization measurements (Aleksandar Cikota, private comm.), although we note that the results presented in the next section are not qualitatively changed if we apply one. We also did not apply a polarization flat field correction, as none were available for dates close to those of our observations. \citet{gonzalez20} finds the effects of applying this correction are negligible, so this should not affect our results. Finally, we remove cosmic rays in each frame using the algorithm of \citet{vandokkum01} through the Astro-SCRAPPY tool \citep{mccully18}.

\section{Analysis}\label{sec:analysis}

To estimate the linear degree of polarization of W0116--0505, as well as of the standard stars, we follow the procedure outlined in \citet{gonzalez20} and the FORS2 User Manual. We first determine the Stokes $Q$ and $U$ parameters as
\begin{eqnarray}\label{eq:stokes}
    Q &=& \frac{2}{N}\ \sum_{i=1}^{N} F(\theta_i)\ \cos(4\theta_i)\nonumber\\
    U &=& \frac{2}{N}\ \sum_{i=1}^{N} F(\theta_i)\ \sin(4\theta_i), 
\end{eqnarray}
\noindent where $\theta_i$ are the retarder plate angles of 0, 22.5, 45 and 67.5 degrees and $N=4$ is the total number of angles used. $F(\theta_i)$ is defined as
\begin{equation}\label{eq:F_th}
    F(\theta_i) = \frac{f_{o,i}-f_{e,i}}{f_{o,i}+f_{e,i}},
\end{equation}
\noindent where $f_{o,i}$ and $f_{e,i}$ are the summed instrumental $o$- and $e$-beam fluxes across all observations taken with retarder plate angle $\theta_i$. Note that $Q$ and $U$ are already normalized by the Stokes parameter $I$. We use this notation throughout the article. To measure each flux we perform aperture photometry using the {\tt{Python}} package {\tt{Photutils}} \citep{bradley19}.  We use 2\arcsec\ diameter apertures and estimate the background contribution using an annulus with inner and outer edge radii of 4\arcsec\ and 7\arcsec, respectively. We note that using a smaller (larger) aperture diameter of 1\arcsec\ (3\arcsec) nominally improves (worsens) the $S/N$ of the detection, with only statistically consistent differences in the polarization amplitude and angle. However, since in the observations of the final OB have seeing comparable to 1\arcsec\ we conservatively chose to use the larger 2\arcsec\ diameter aperture to minimize aperture losses between different plate angles due to seeing variations during those observations that could potentially induce systematic uncertainties in the $Q$ and $U$ parameter estimates. We analyze all OBs with the same aperture size for consistency. 

For each source we then remove the instrumental background polarization at its position within the field of view (FoV) using the corrections for $Q$ and $U$ provided by equation (12) and Table 4 of \citet{gonzalez20}.  These corrections are larger for objects towards the edges of the FoV, but typically negligible near the center, where W0116--0505 and the standard stars were placed. Note that our conclusions are qualitatively unaffected if we do not apply these corrections. 

We calculate the linear degree polarization $p$ as
\begin{equation}\label{eq:pol_frac}
    p = \sqrt{Q^2 + U^2}, 
\end{equation}
\noindent and the polarization angle $\chi$ as
\begin{equation}
    \chi = \frac{1}{2}\ \arctan\left(\frac{U}{Q}\right).
\end{equation}
\noindent Finally, we correct the angle $\chi$ for chromatism of the half wave plate by subtracting the zero angle estimated for the $R$-band of --1.19~deg in the FORS2 User Manual (see Table 4.7). Table \ref{tab:pol_tab} shows the linear polarizations and polarization angles measured for W0116--0505 and the standard stars.

\begin{deluxetable}{lccc}

    \tablecaption{Linear Polarization Measurements\label{tab:pol_tab}}
    
    \tablehead{
        \colhead{Target} &
        \colhead{OB} &
        \colhead{$P$} & 
        \colhead{$\chi$} \\
        \colhead{} &
        \colhead{} &
        \colhead{(\%)} &
        \colhead{($^{\circ}$)}
    }
    
    \tabletypesize{\small}
    \tablewidth{0pt}
    \tablecolumns{4}
    
    \startdata
    W0116--0505  & \phn\phn2886768 & 11.5$\pm$3.5 & 72$\pm$26\\
                 & \phn\phn2886765 & 11.1$\pm$3.7 & 75$\pm$34\\
                 & \phn\phn2886772 & 10.2$\pm$3.4 & 74$\pm$35\\
                 & \phn\phn2886622 & 10.6$\pm$4.1 & 77$\pm$55\\
                 & {\bf{Combined}} & {\bf{10.8$\pm$1.9}} & {\bf{74$\pm$\phn9}}\\
    \\
    \multicolumn{4}{l}{{\it{Polarization Standard}}}\\
    BD--12 5133  &       200277970 & 4.12$\pm$0.05 & 146.5$\pm$0.3\\
    \\
    \multicolumn{4}{l}{{\it{Zero-polarization Standards}}}\\
    WD 0310--688 &       200277988 & 0.04$\pm$0.05 & \nodata\\
    WD 2039--202 &       200278006 & 0.12$\pm$0.09 & \nodata\\
    \enddata
    
\end{deluxetable}

To estimate the uncertainties in $p$ and in $\chi$, we first estimate the nominal flux uncertainties for each measurement of $f_{o,i}$ and $f_{e,i}$. Then, assuming the noise is Gaussian, we produce 1,000 simulated values of $f_{o,i}$ and $f_{e,i}$ for each observation of each source, and then combine them to form 1,000 pairs of $Q$ and $U$ for each source. We then estimate uncertainties in $P$ and $\chi$ as the dispersion observed for the simulated pairs of $Q$ and $U$. 

For the zero-polarization standards, WD 0310--688 and WD 2039--202, we obtain linear polarizations of 0.04$\pm$0.05\% and 0.12$\pm$0.09\%. As expected, both are consistent with no polarization. For the polarization standard, BD--12 5133, we find a linear polarization of 4.12$\pm$0.05\%, and a polarization angle of 146.5$\pm$0.3~deg. These values are consistent at the 2$\sigma$ level with the linear polarization fraction and angle of 4.02$\pm$0.02\% and 146.97$\pm$0.13~deg measured by \citet{fossati07} using the FORS1 instrument and the $R_{\rm Bessel}$ band. The slightly higher polarization is likely due to the differences between the $R_{\rm Special}$ filter used in our observations with the $R_{\rm Bessel}$ filter used by \citet{fossati07}. Figure \ref{fg:spec_and_filters} shows the two filter curves\footnote{Filter curves were obtained from \url{https://www.eso.org/sci/facilities/paranal/instruments/fors/inst/Filters/curves.html}}. Since $R_{\rm Bessel}$ extends somewhat redder than $R_{\rm Special}$ and \citet{fossati07} measures a polarization that increases mildly towards 4.37$\pm$0.05\% at $V$-band, a slightly higher degree of polarization in the $R_{\rm Special}$ band might be expected. 

\begin{figure}
  \includegraphics[width=0.475\textwidth]{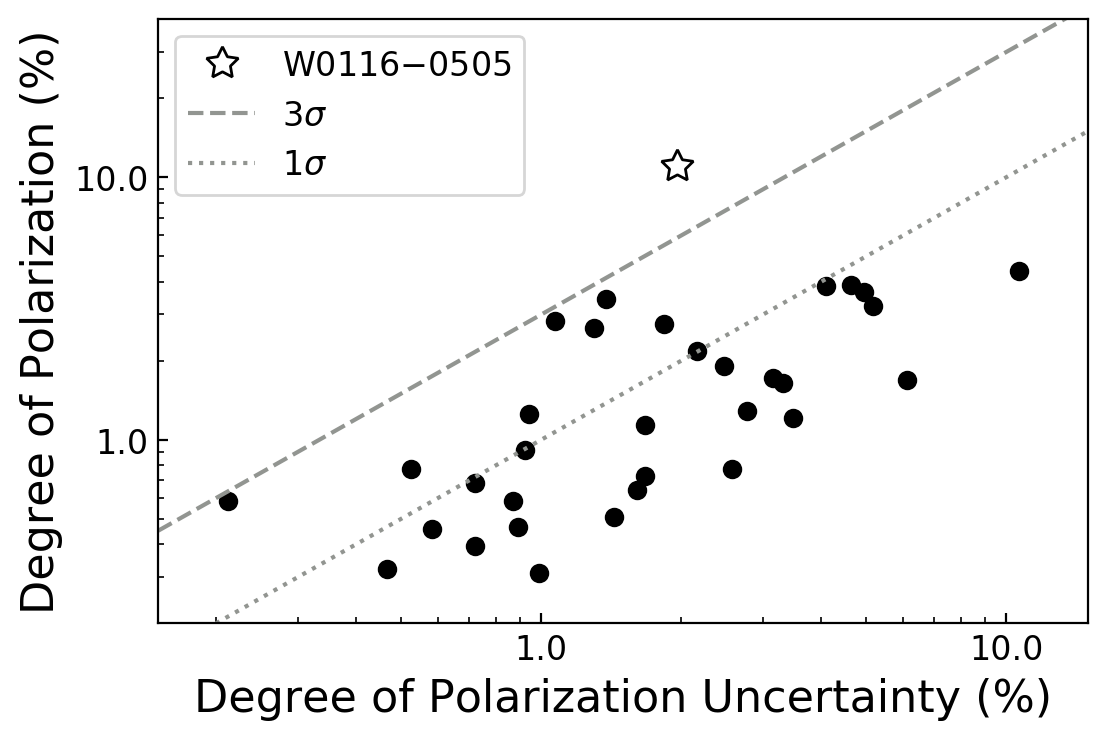}
  \caption{Degree of polarization and its uncertainty for W0116--0505 compared to that of all other targets detected with $S/N>5$ $e$-beam fluxes in single observations without obvious systematic problems. The diagonal gray lines show linear polarizations detected at 1$\sigma$ (dotted) and 3$\sigma$ (dashed). W0116--0505 is the only source in the field with a linear polarization detection of $>3\sigma$.}
  \label{fg:pol_frac}
\end{figure}

Following the procedure described above we measure for W0116--0505 a polarization of 10.8$\pm$1.9\% and a polarization angle of 74$\pm$9~deg when combining the observations of all four OBs. If we instead analyze the observations separately for each OB, we find consistent values for both the polarization fraction and angle, albeit with larger uncertainties, as shown in Table \ref{tab:pol_tab}. This shows that the high linear polarization detected for W0116--0505 is not driven by outliers in the observations. As a further check, we follow the procedure above to estimate the linear polarization for all objects in the FoV of W0116--0505 detected with $S/N\ge 5$ in the e-beam of a single exposure that were not affected by obvious issues such as artifacts and blending with neighboring sources. Figure \ref{fg:pol_frac} shows the linear polarizations detected and their uncertainties for all of these sources, and W0116--0505 is the only source that has a detection above 3$\sigma$. This confirms that the measurement is not driven by a problem with data reduction or the analysis procedures. Furthermore, it shows that the detected polarization is not caused by interstellar gas and dust. As a further check to constrain the effects of interstellar polarization, we take all 55 stars in the catalog of \citet{heiles00} of stellar polarizations that are within 10~deg of W0116--0505 and combine their polarization signals by adding their $Q$ and $U$ Stokes parameters assuming the same value of $I$. As their intrinsic polarization will have random orientations, the only coherent term that should survive the averaging is that caused by the ISM. This estimate yields $p_{\rm interstellar}=0.064$\%, well below the linear polarization fraction detected for W0116--0505. In the next section we discuss the implication of this high degree of polarization detected for W0116--0505.  

\section{Discussion}\label{sec:discussion}

\subsection{The Case for Scattered Light in BHDs}

As discussed in \S\ref{sec:intro}, \citet{assef16} identified a group of 8 BHDs by studying the SEDs of a sample of Hot DOGs. When accounting for selection biases, \citet{assef16} found that up to 12\% of Hot DOGs could be classified as BHDs, although suggested the true fraction is likely lower. Considering the results of their SED modeling, \citet{assef16} determined three possible sources of the blue-excess emission of BHDs: a) a secondary AGN in the system; b) a luminous unobscured starburst outside the range of the templates of \citet{assef10}; or c) scattered light from the highly-obscured central engine. 

\citet{assef16} and \citet{assef20} conducted detailed studies using {\it{HST}}/WFC3 and {\it{Chandra}}/ACIS observations as well as ground-based photometry and spectroscopy of three BHDs, W0204--0506, W0220+0137 and W0116--0505 (the subject of the present polarimetry study), in the context of each hypothesis. Based on the lack of soft X-ray emission detected in all three sources, and on the centrally concentrated undisturbed morphologies of W0220+0137 and W0116--0505, the presence of an additional, unobscured AGN in the system (hypothesis a) was judged to be highly unlikely. They also determined that an unobscured star-burst (hypothesis b) was unlikely to be the dominant source of the rest-frame UV/optical emission due to the high SFRs required, but could not rule this out completely (see \citealt{assef20} for details). For W0204--0506, \citet{assef20} pointed out that the disturbed morphology with some significantly extended UV flux implied that SF could account for a fraction of the observed blue-light excess. Both studies, based on the challenges of the other scenarios, concluded that the most likely source of the excess blue light emission is scattered emission from the highly-obscured central engine (hypothesis c). 

Further insights can be obtained by analyzing the variability of the UV continuum, as no significant variability would be expected for an unobscured starburst. For W0116--0505, we can compare the $r$-band fluxes of SDSS and PanSTARRS PS1 to cover a significantly long timescale, particularly since there are almost no color effects between the surveys in this band \citep{tonry12}. We find that the PSF-fitting magnitudes are fainter by about 0.14$\pm$0.06~mag in the PS1 DR2 mean object catalog as compared to SDSS DR16. The PS1 observations were obtained between 60 and 513 rest-frame days after the SDSS observations. Variability supports hypotheses a and c, in which the rest-frame UV flux is dominated by AGN emission. This amplitude, while small, is somewhat above what would be expected for a luminous AGN in such a timescale \citep[e.g.,][]{vandenberk04, macleod10}, and it could be due to a systematic difference between the PS1 and SDSS photometry of the source. A deeper analysis, beyond the scope of this study, would be needed to address systematic uncertainties.

As discussed in \S\ref{sec:intro}, polarization can be a powerful tool to confirm that the source of the blue excess is indeed scattered AGN light, as we would not expect significant linear polarization from star-formation. The linear polarization of $p = 10.8\pm 1.9\%$ found for W0116--0505 in the $R_{\rm Special}$ band in \S\ref{sec:analysis} shows that the UV emission is dominated by scattered emission from the central engine rather than star-formation, confirming the conclusions of \citet{assef20} for this source. In the following sections we discuss the effects that differing polarization between the UV continuum and the broad emission lines could have on our broad-band polarization measurement, as well as the implications for obscuration geometry and for the properties of the scattering medium in W0116--0505.

\subsection{Polarized Light of Continuum vs. Emission Lines}

The broad emission lines could in principle show different linear polarization properties from the continuum as they arise from physically distinct structures, namely the broad-line region and the accretion disk. As discussed in \citet{assef20}, W0220+0137 is classified as an Extremely Red Quasar (ERQ) and W0116--0505 is classified as ERQ-like \citep[see][]{hamann17,goulding18}. \citet{alexandroff18} found for three objects, an ERQ, an ERQ-like and a type 2 quasar, a 90 deg difference between the polarization angle of the Ly$\alpha$, N{\sc v} and C{\sc iv} broad emission lines in certain velocity ranges and the continuum emission, with a linear degree of polarization of $\sim 10\%$ in the continuum and $\sim 5\%$ in the emission lines. Our broad-band encompasses both UV continuum and the C{\sc iv} broad emission line, which has an equivalent width of 86$\pm$17\AA. Hence, the linear degree of polarization of the continuum alone and/or the C{\sc iv} line may separately exceed the 10.8$\pm$1.9\% determined in \S\ref{sec:analysis}. If we assume that both the continuum and the emission line fluxes have the same degree of polarization, $p_{\rm int}$, but with polarization angles differing by 90 degrees, the measured $p$ relates to $p_{\rm int}$ by
\begin{equation}
    p_{\rm int}\ =\ \frac{p}{|f_c - f_l|},
\end{equation}
\noindent where $f_c$ and $f_l$ are the fraction of the integrated flux in the $R_{\rm Special}$ band coming from the continuum and the emission line respectively. Modeling the continuum as a linear function of the wavelength and the emission line as a single Gaussian, as outlined in \citet{assef20}, we find that 82\% of the flux in the $R_{\rm Special}$ band is contributed by the continuum, implying $p_{\rm int}=17$\%. Future spectropolarimetric observations or imaging polarimetry observations in multiple broad-bands could further elucidate this point. We direct the reader to \citet{alexandroff18} for a more comprehensive overview of possible geometries that can potentially explain the high degree of polarization and a swing in the polarization angle between the continuum and the broad emission lines. 

\subsection{The Scattering Medium of W0116--0505}\label{ssec:medium}

\citet{assef20} argued that if the UV was dominated by scattered light, the scattering medium was more likely ionized gas rather than dust, as there was no significant ``bluening'' of the underlying AGN spectrum according to the AGN template of \citet{assef10}. However, they could not discard dust as the scattering medium as there were too few photometric data points to rule out the possibility of ISM dust reddening canceling the bluening of the input spectrum. Furthermore, for some types of dust no bluening may be observed. For example, \citet{zubko00} show that for the type of dust proposed by \citet{mathis77}, no bluening is observed at $\lambda\lesssim 2500$\AA\ for scattering angles $\gtrsim~30~\rm deg$.

\begin{figure*}
  \plotone{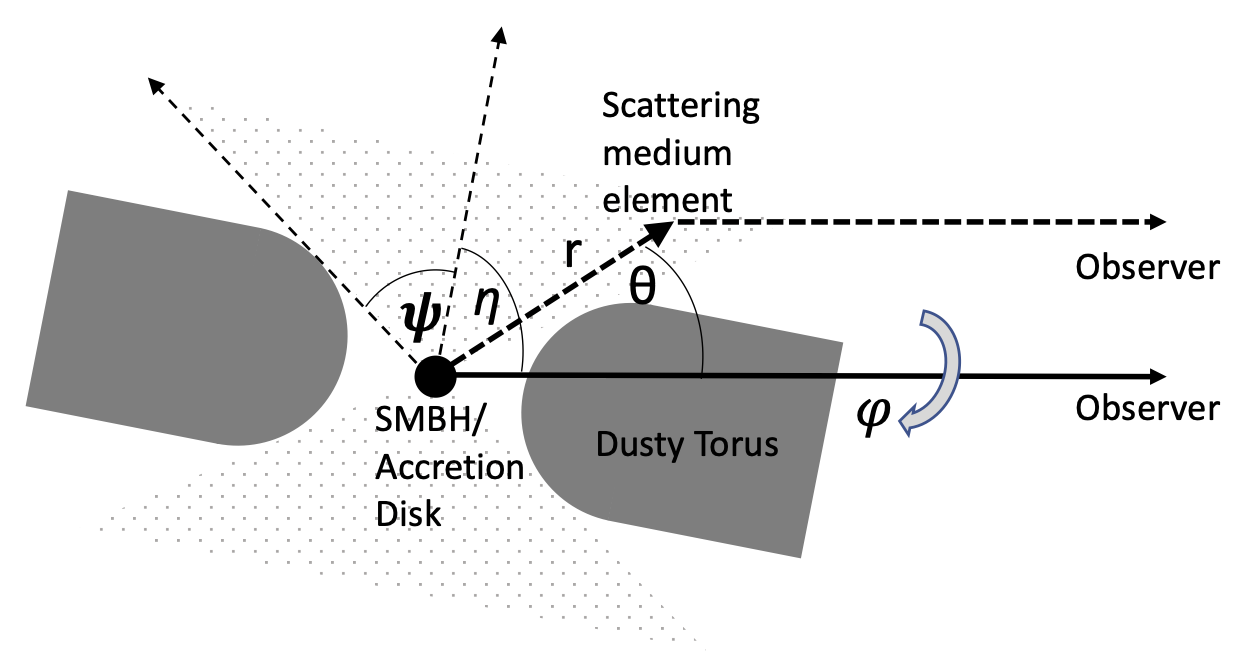}
  \caption{Schematic drawing of the geometry for the optically-thin scattering model discussed in \S\ref{ssec:medium}.}
  \label{fg:cartoon}
\end{figure*}

In order to analyze the properties of the scattering medium, we consider the degree of polarization detected and the total scattered flux in the context of a simple model, similar to that explored by \citet{miller91}. Furthermore, we take into consideration the fact that the scattered continuum is unreddened and that the UV continuum is resolved in the {\it{HST}}/WFC3 F555W images \citep[see][and discussion below for details]{assef20}. To construct the model, we first assume that the scattering medium is optically thin to the scattered radiation. In other words, we assume that the UV photons reaching us have only experienced a single scattering event along their path. We also assume that the accretion disk is covered by a traditional torus with two polar openings, each of angular half-size $\psi$, through which UV light can escape and interact with the scattering medium. We study the geometry of the system in spherical coordinates with the polar axis pointing at our line of sight, as shown in Figure \ref{fg:cartoon}. In these coordinates, $r$ corresponds to the distance of a scattering particle from the SMBH. Conveniently, the scattering angle of that scattered radiation corresponds to the particle's polar angle, so we refer to both as $\theta$ in this section. The polarization angle of the radiation scattered by that particle, $\chi_{\rm particle}$, is conveniently related to its azimuthal angle $\phi$ by $\chi_{\rm particle} = \phi + \pi/2$. In this model the centers of the openings in the torus are respectively inclined by polar angles $\theta=\eta$ and $\theta=\eta_S+\pi$ from the line of sight. 

In general, we can say that the total flux from the AGN accretion disk scattered into our line of sight is given by
\begin{equation}\label{eq:tot_scatt_flux}
    F_{\rm Scatt}\ =\ \int \frac{F_{\rm AD}}{r^2}\ \frac{d\sigma}{d\Omega}(\theta)\ n_{\rm H}\ dV,
\end{equation}
\noindent where $F_{\rm AD}$ is the flux we would receive if we had an unobscured line of sight to the accretion disk, $\frac{d\sigma}{d\Omega}(\theta)$ is the differential scattering cross section per hydrogen nucleon for scattering angle $\theta$ and $n_{\rm H}$ is the number density of hydrogen atoms. The integral is carried over the volume of all the regions of the galaxy illuminated by the AGN through the openings in the torus. We further assume that $n_{\rm H}$ is only a function of $r$, and note that the column density $N_{\rm H}$ is given by
\begin{equation}\label{eq:column_density}
    N_{\rm H}\ =\ \int n_{\rm H}(r) dr .
\end{equation}

Equation (\ref{eq:tot_scatt_flux}) can be re-written to depend on the fraction of the accretion disk flux scattered into our line of sight, $f_{\rm Scatt} = F_{\rm Scatt}/F_{\rm AD}$. We, however, do not have a direct measurement of $F_{\rm AD}$, as we can only probe the brightness of the torus from the IR emission due to the dust obscuration. The relation between the accretion disk luminosity and the torus luminosity is given by the covering factor of the torus, which for our model of a torus with two polar openings of angular half-size $\psi$ is given by $\cos \psi$. In other words, 
\begin{equation}\label{eq:Lir_Lad}
    L_{\rm IR}\ =\ \cos\psi~ L_{\rm AD}.
\end{equation}
\noindent \citet{assef20} found that the AGN that reproduced the scattered component in the UV and optical had $\epsilon = 0.6\%$ of the luminosity of the obscured AGN that reproduced the mid-IR component when both components were modeled with the same AGN template from \citet{assef10} (see Fig. \ref{fg:sed}). We assume that the coverage fraction for the \citet{assef10} empirical AGN SED template used is $1/2$, since \citet{assef13} found that on average $\sim$50\% of quasars appeared obscured in a sample similar to that used by \citet{assef10} to define their template. We also note that \citet{roseboom13} found a dust covering fraction of $\sim$40\% for quasars in a slightly higher luminosity range to those used by \citet{assef10}. Considering this in conjunction with equation (\ref{eq:Lir_Lad}), we find that 
\begin{equation}\label{eq:ref_scatt}
    f_{\rm Scatt}\ =\ 2 \epsilon \cos\psi.
\end{equation}
\noindent Using equations (\ref{eq:column_density}) and (\ref{eq:ref_scatt}), we can rewrite equation (\ref{eq:tot_scatt_flux}) as
\begin{equation}\label{eq:eq_model_1}
    2 \epsilon \cos\psi\ =\ N_{\rm H}\ S_1,
\end{equation}
\noindent where $N_{\rm H}$ is the hydrogen column density and 
\begin{equation}
    S_1\ =\ \int \frac{d\sigma}{d\Omega}(\theta)\ d\Omega .
\end{equation}
\noindent We note that while equation (\ref{eq:Lir_Lad}) could potentially be used to constrain the maximum size of $\psi$ so as to avoid an accretion disk luminosity that would exceed the most luminous type 1 quasars known \citep[e.g.,][]{schindler19}, these constraints are weaker than the ones discussed below.

Using the same assumptions and geometry we also calculate the total degree of polarization and polarization angle by noting that the Stokes parameters $Q$ and $U$ are additive. Furthermore, we orient the zero point of the azimuthal axis such that the torus openings are found along $\phi=0,\pi$, which is equivalent to saying that the overall polarization angle of the system will be $\pi/2$. In this case, $U$ is always equal to zero, and the degree of polarization is given by $p\ =\ -Q$. We can rewrite the latter expression as:
\begin{equation}\label{eq:eq_model_2}
   2\ p\ \epsilon\ \cos\psi\ =\ -N_{\rm H}\ S_2,
\end{equation}
\noindent where 
\begin{equation}
    S_2\ =\ \int \frac{d\sigma}{d\Omega}(\theta)\ p_{\rm particle}(\theta)\ \cos(2\phi+\pi)\ d\Omega,
\end{equation}
and $p_{\rm particle}(\theta)$ is the polarization degree caused by the scattering particle when irradiated by unpolarized light. We note that by dividing equations (\ref{eq:eq_model_1}) and (\ref{eq:eq_model_2}), we can write the overall degree of polarization as
\begin{equation}\label{eq:eq_model_3}
    p\ =\ \frac{-S_2}{S_1}.
\end{equation}

\subsubsection{Thomson Scattering}

\begin{figure*}
  \includegraphics[width=\textwidth]{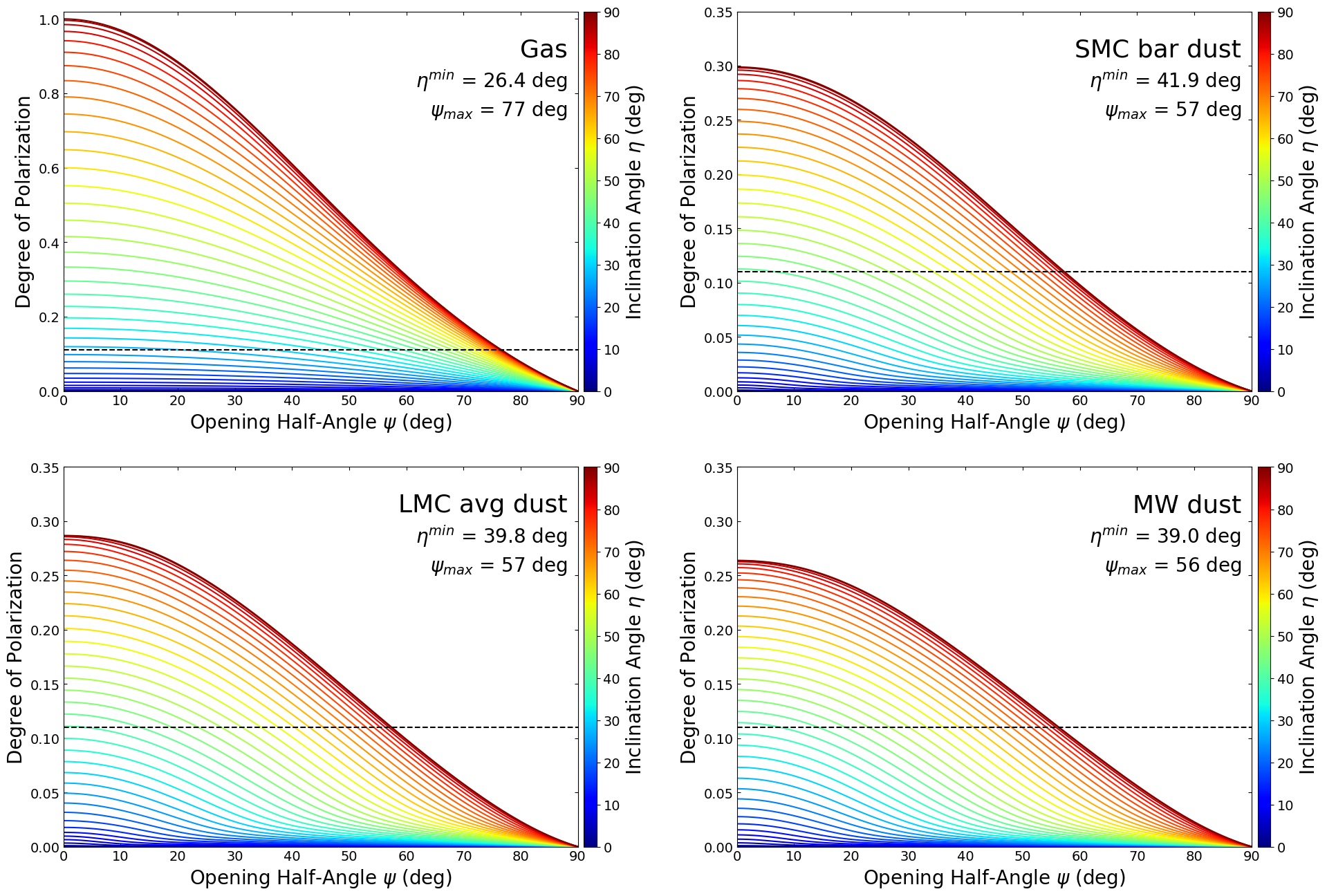}
  \caption{Degree of polarization expected as a function of opening angle $\psi$ and inclination angle $\eta$ of the dust torus for the model discussed in \S\ref{ssec:medium}, for purely ionized gas (top left), and the SMC bar dust mixture (top right), LMC average dust mixture (bottom left) and MW dust mixture (bottom right) of \citet{draine03}. The inclination angle $\eta$ is indicated by the line color and the color bar on the right. The horizontal dashed line shows the measured polarization fraction for W0116--0505.}
  \label{fg:pol_frac_model}
\end{figure*}

First, let's consider the simplest scenario, where the scattering medium is purely composed of ionized hydrogen. In this case we have simply Thomson scattering due to free electrons, with the density of free electrons being the same as the density of hydrogen nucleons. The differential cross section per hydrogen nucleon is then given by
\begin{equation}
    \frac{d\sigma}{d\Omega}(\theta)\ =\ \frac{1}{2}\ r_0^2\ (1+\cos^2\theta),
\end{equation}
\noindent where $r_0$ is the classical electron radius, and the polarization degree per particle is given by
\begin{equation}
    p_{\rm particle}\ =\ \frac{1-\cos^2\theta}{1+\cos^2\theta} .
\end{equation}
The top left panel of Figure \ref{fg:pol_frac_model} shows the degree of polarization as a function of $\psi$ and $\eta$. The minimum inclination angle for the openings is $\eta^{\rm min}=26.4~\rm deg$, as lower inclination angles cannot reproduce the observed degree of polarization of $p=10.8\%$. At that inclination angle, $\psi$ would have to be close to zero in order to not depolarize the combined emission. As $\eta$ becomes larger, $\psi$ becomes larger as well in order to depolarize the combined emission and bring it down to the observed value. When the opening is inclined perpendicularly to the line of sight (i.e., the torus is edge-on), it reaches its maximum possible size of $\psi_{\rm max} = 77~\rm deg$. 

\begin{figure*}
  \plotone{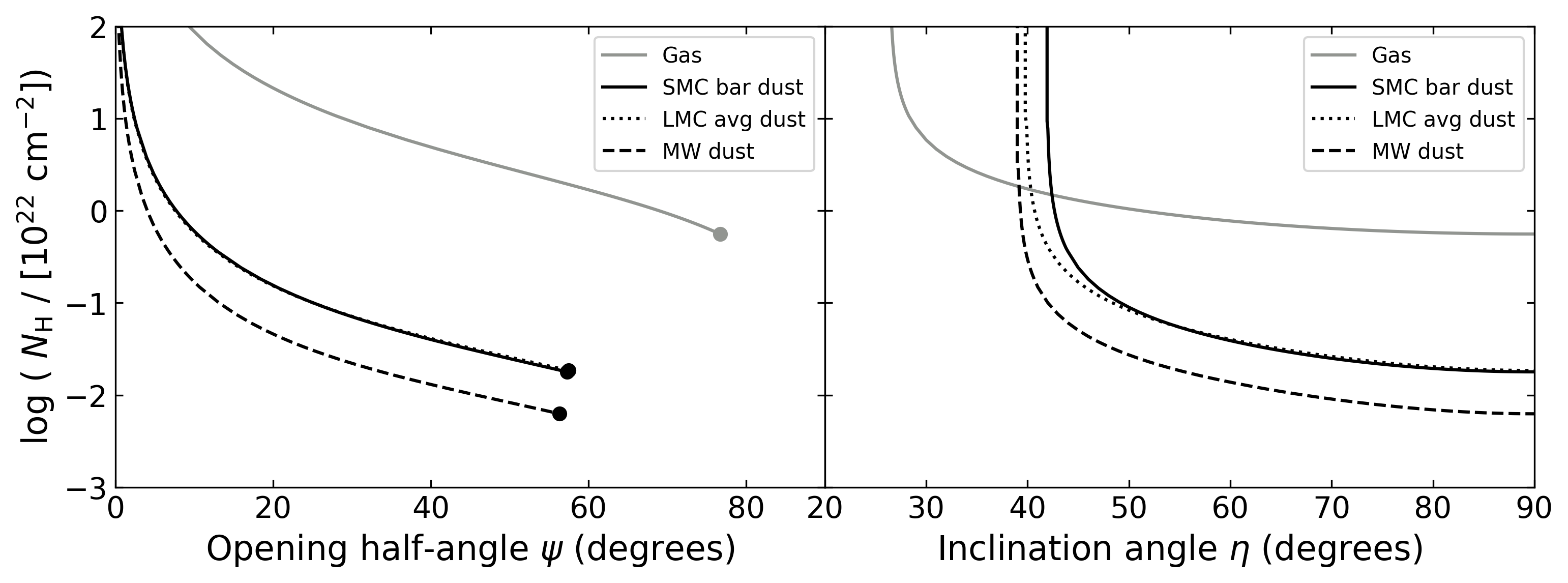}
  \caption{Constraints on the scattering medium $N_{\rm H}$ as a function of the opening angle $\psi$ (left) and the inclination angle $\eta$ (right) of the dust torus for the model discussed in \S\ref{ssec:medium}. Note that in the left panel the lines from the SMC bar and LMC average dust mixtures almost completely overlap.}
  \label{fg:NH_model}
\end{figure*}

In addition to analyzing the degree of polarization expected for a given geometry, we can also use equations (\ref{eq:eq_model_1}) and (\ref{eq:eq_model_2}) to find for each value of $\eta$ the values of $\psi$ and $N_{\rm H}$ that simultaneously reproduce the observed degree of polarization and the fraction of the incident flux scattered into our line of sight. Figure \ref{fg:NH_model} shows the required value of $N_{\rm H}$ as a function of $\psi$ and $\eta$. For all geometries we would require column densities of $N_{\rm H}>5\times 10^{21}~\rm cm^{-2}$. The much larger values required for low inclination angles and small torus openings potentially conflict with the model assumption of an optically thin medium from the line of sight of the observer. If we consider that multiply scattered photons lose the ensemble polarization properties, then an optically thick medium would require a larger minimum inclination angle in order to reproduce the observed degree of polarization. 

\begin{figure}
  \includegraphics[width=0.475\textwidth]{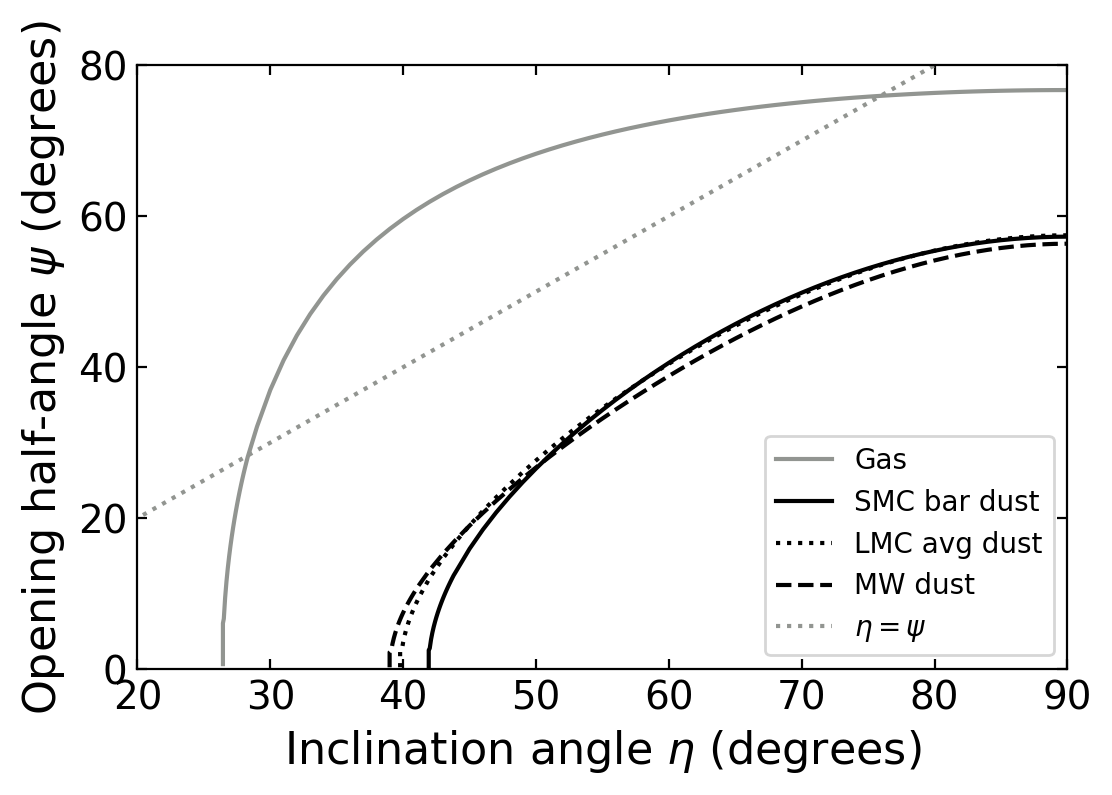}
  \caption{Opening angle $\psi$ as a function of the inclination angle $\eta$ of the dust torus for the model discussed in \S\ref{ssec:medium}. Line styles have the same meaning in Fig. \ref{fg:NH_model}.}
  \label{fg:psi_th_model}
\end{figure}

Since dust particles have a much larger scattering cross-section than free electrons, their contribution to scattering can easily dominate over that of Thomson scattering off free electrons. \citet{draine03} conducted a detailed study of the scattering properties of UV and optical light for different dust-to-gas mixtures, and found the dust differential scattering cross-section per hydrogen nucleon in the UV for the type of dusty ISM typically found in the SMC, LMC and Milky Way to be between $\sim 5\times 10^{-24}$ and $\sim 10^{-21}~\rm cm^2~\rm H^{-1}~\rm sr^{-1}$, depending on the exact dust mixture and scattering angle. In a fully ionized hydrogen medium, the differential scattering cross section per hydrogen nucleon would be of order $r_0^2~\rm H^{-1}~\rm sr^{-1} = 8\times 10^{-26}~\rm cm^{2}~\rm H^{-1}~\rm sr^{-1}$. Hence, dust scattering from the former should dominate over free electrons unless the scattering medium is very dust-deficient. We expect dust to be present in significant quantities in the ISM of the host galaxy as, by selection, Hot DOGs have bright dust emission in the mid-IR and are typically well detected in the far-IR \citep{jones14, wu14, diaz-santos21}\footnote{While \citet{diaz-santos21} did not find extended dust emission in ALMA band 8 observations of W0116--0505, their observations only had a spatial resolution of $\sim$2~kpc.}. Hence, the Thomson scattering scenario would only be plausible if the scattering medium is within the dust sublimation radius of the accretion disk, which we estimate to be 8.7~pc using equation (1) of \citet{nenkova08} and the bolometric luminosity of W0116--0505 measured by \citet{tsai15}, adjusted for the differences in the cosmological model assumed. In order to have a direct line of sight towards the inner regions of the torus but without a direct line of sight to the accretion disk, we need $\psi\sim \eta$, which is achieved at two different inclination angles of $\eta\approx 28~\rm deg$ and $\approx 76~\rm deg$ as shown in Figure \ref{fg:psi_th_model}. 

The extension of the UV emission in the {\it{HST}}/F555W image, which has a half light radius of 0.9~kpc as mentioned in \S\ref{ssec:prev_data}, is also hard to reconcile with the idea of the scattering occurring within the sublimation radius, which is two orders of magnitude smaller. UV emission in extended scales would either require substantial scattering by the ISM, which is likely dominated by dust, or a significant SF component, which was deemed unlikely by \citet{assef20}. Furthermore, the host galaxy emission is more extended in a direction that is close to perpendicular to the polarization angle, as shown in Figure \ref{fg:HST}, suggesting this extended component is along the regions illuminated by AGN and consistent with being scattered light.

Considering all of this, we conclude it is unlikely that Thomson scattering can be the dominant scattering mechanism in W0116--0505, and so substantial dust scattering is very likely. We now look at optically thin dust in the ISM as the possible scattering medium.

\subsubsection{Scattering by Optically-Thin Dust}\label{sssec:thin_dust_scattering}

We conduct the same analysis done for Thomson scattering but using the differential cross section per hydrogen nucleon and degree of polarization provided by \citet{draine03} for an $R_{\rm V}=3.1$ Milky-Way (MW) dust mixture, as well for the dust mixtures for the SMC bar and  LMC average of \citet{weingartner01}\footnote{The differential cross sections and degree of polarization curves were retrieved from \url{http://www.astro.princeton.edu/~draine/dust/scat.html}}. Specifically, we use those determined for a rest-frame wavelength of 1600~\AA, which is close in the rest-frame wavelength that corresponds to the effective wavelength of the $R_{\rm Special}$ filter. Figure \ref{fg:pol_frac_model} shows the expected observed polarization for these dust mixtures as a function of $\eta$ and $\psi$. The SMC bar dust mixture provides the highest possible linear polarization, but all three are easily capable of reproducing the observed value, with minimum inclination angles of $\eta^{\rm min} = 41.9$, 39.8 and 39.0~deg for the SMC, LMC and MW dust mixtures, respectively. At an inclination of 90 deg, the opening must be again maximal to allow for enough depolarization to match the observed degree of polarization. The maximum openings are somewhat smaller than for the Thomson scattering case, with $\psi_{\rm Max} = 57$, 57 and 56~deg for the three dust mixtures. Figure \ref{fg:NH_model} shows the required values of $N_{\rm H}$ as a function of $\psi$ and $\eta$ to reproduce the observations. We find that the values are very similar for the SMC and LMC dust mixtures, both requiring $N_{\rm H}>1.8\times 10^{20}~\rm cm^{-2}$, while the MW dust mixture only requires $N_{\rm H} > 6\times 10^{19}~\rm cm^{-2}$. 

The hydrogen column densities expected in the case of optically thin dust mixtures are consistent with the lack of reddening found by \citet{assef20} for the scattered light ($E(B-V)<0.02~\rm mag$, 1$\sigma$ limit). \citet{maiolino01} studied the dust-to-gas ratio in the circumnuclear regions of nearby AGN, and found ratios that were significantly below the Galactic standard value of $A_V/N_{\rm H} = 4.5\times 10^{-22}~\rm mag~\rm cm^2$ \citep{guver09}, or $E(B-V)/N_{\rm H} = 1.5\times 10^{-22}~\rm mag~\rm cm^2$ assuming $R_V=3.1$. \citet{maiolino01} found a median ratio in nearby AGN of $E(B-V)/N_{\rm H} = 1.5\times 10^{-23}~\rm mag~\rm cm^{2}$, about 10 times lower than the Galactic standard albeit with a large variance. A reddening of $E(B-V)<0.02~\rm mag$ would then require $N_{\rm H}<1.3\times 10^{21}~\rm cm^{-2}$ for the \citet{maiolino01} median value, which is consistent with the results of our simple model for $\eta\gtrsim 40-50~\rm deg$ and $\psi\gtrsim 10-25~\rm deg$, depending on the exact dust mixture. For the Galactic standard value we would instead need $N_{\rm H}<1.3\times 10^{20}~\rm cm^{-2}$, which is only achievable for the MW dust mixture for $\eta\gtrsim 60~\rm deg$ and $\psi\gtrsim 40~\rm deg$.

A more complex model could consider multiple additional openings in the torus through which the light from the central engine is escaping into the ISM. In the case of multiple openings in the dust distribution, each will create linearly polarized emission but with different polarization degrees and angles, resulting in dilution of the combined linear polarization. Such a scenario would require a higher inclination of the main openings through which the light is escaping in order to achieve the same overall polarization of 10.8\%. A contribution from star-formation would also have the effect of diluting the linear polarization signal, although \citet{assef20} found it unlikely that there is a significant contribution from star-formation to the UV/optical SED of W0116--0505, as mentioned earlier. 

\subsection{Outflow in W0116--0505 as a Possible Scattering Medium}\label{ssec:outflow}

In the previous section we analyzed the observed degree of polarization in W0116--0505 in the context of a simple model in which an AGN is surrounded by a torus with bi-polar openings through which light can escape and interact with a uniform ISM. We showed that for reasonable dust mixtures, the model is able to reproduce three key aspects of our observations, namely the degree of polarization of 10.8\%, a fraction of the total flux scattered into our line of sight  of 0.6\%, and that the scattered flux is consistent with no dust reddening. While the model is quite simple and depends only on three parameters ($N_{\rm H}$, $\psi$ and $\eta$), the parameters are heavily degenerate, suggesting a more complicated model is not warranted without further constraints. 

While the observations are consistent with a regular ISM being the source of the scattering, this is clearly not a unique scenario. In particular, the more sophisticated model proposed by \citet{alexandroff18} should be able to explain the observations as well, as the objects analyzed in that work share a number of similarities with W0116--0505. In that model the scattering medium consists of a fast outflow escaping through a polar opening in the AGN torus. Here we consider as well the possibility of an outflow being responsible for the scattered light, but with a different approach. \citet{finnerty20} found that the AGN in W0116--0505 is driving a massive gas outflow with an estimated total mass of $M_{\rm gas} = 1.6\times 10^{9}~M_{\odot}$, based on the width and luminosity of its very broad and blue-shifted [O{\sc iii}] emission lines. The luminous [O{\sc iii}] emission implies this outflow is necessarily being illuminated by the AGN, suggesting the outflow could also play a role in the scattering. 

We can make a rough estimate of the hydrogen column density in the outflow by noting that $N_{\rm H} \ge {\langle}n_{\rm H}{\rangle} R_{\rm out}$ for radially declining density profiles such as an isothermal sphere, with the equality corresponding to the limit case of a uniform gas distribution. We take the mean electron density of ${\langle}n_e{\rangle} = 300~\rm cm^{-3}$ and outflow extension of $R_{\rm out}=3~\rm kpc$ assumed by \citet{finnerty20}, and assume that the outflow is primarily composed of ionized hydrogen and helium with the helium number density being 10\% of that of hydrogen \citep[as in][]{carniani15}, implying that ${\langle}n_e{\rangle} = {\langle}n_{\rm H}{\rangle}/1.2$. With these assumptions we estimate $N_{\rm H}\ge 2.3\times 10^{24}~\rm cm^{-2}$, implying a Compton-thick outflow. Using the \citet{maiolino01} median dust-to-gas ratio, this column density would translate to a reddening of $E(B-V)\ge 35~\rm mag$, although it is important to note that this value is of limited accuracy given the substantial uncertainties in the assumed values of ${\langle}n_e{\rangle}$ and $R_{\rm out}$ as discussed by \citet{jun20} as well as in the assumed dust-to-gas ratio.

\begin{figure*}
  \plotone{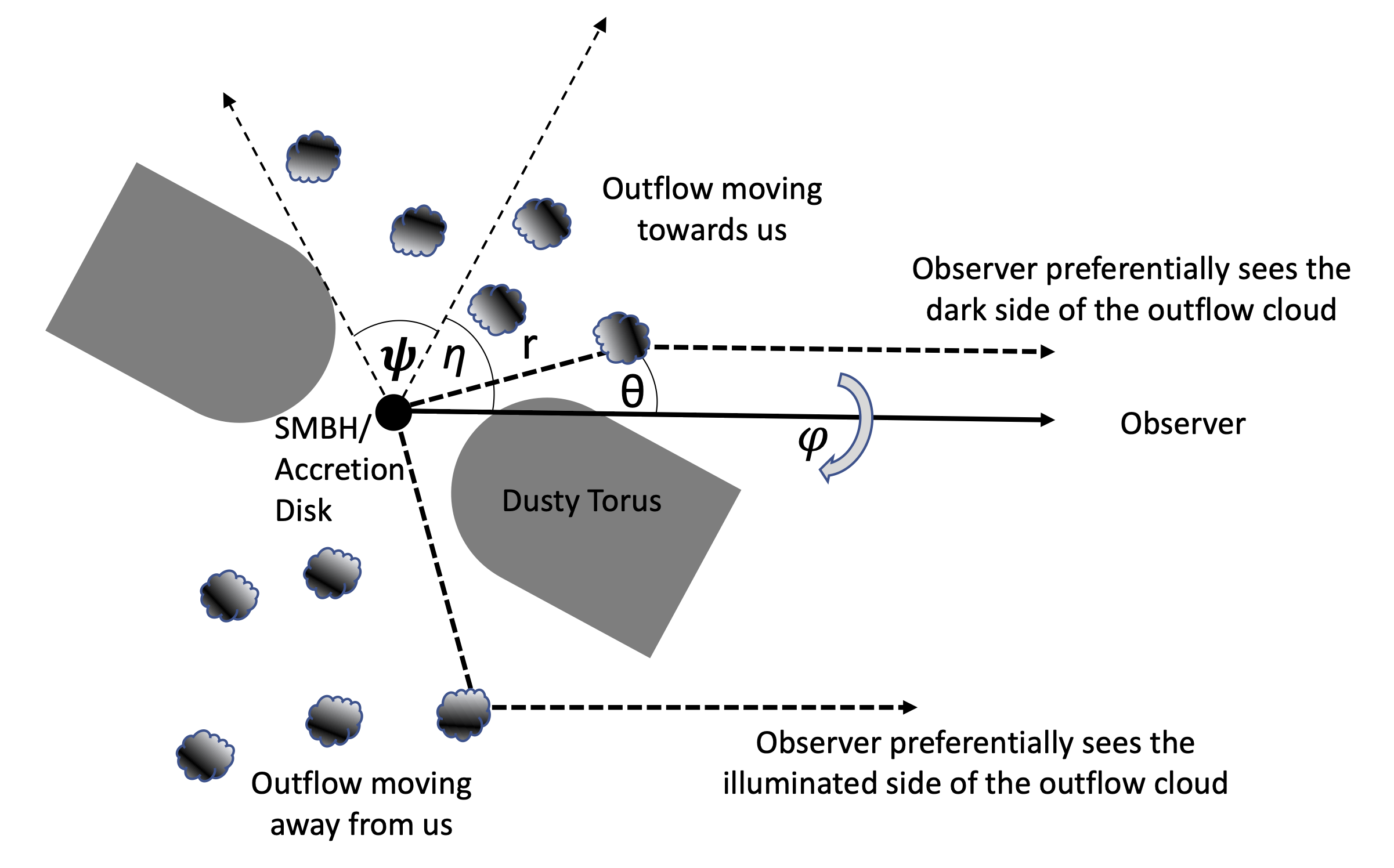}
  \caption{Schematic drawing of the geometry for the optically-thick outflow scattering model discussed in \S\ref{ssec:outflow}.}
  \label{fg:cartoon_outflow}
\end{figure*}

In a more physically motivated picture, it might be possible that the AGN in W0116--0505 is launching polar, possibly dusty outflows as is commonly seen in nearby counterparts, albeit at much lower luminosities and outflow masses \citep[e.g.,][]{schlesinger09, rupke17, stalevski17, alonsoherrero21}, which by themselves might be responsible for the openings in the dust torus. Figure \ref{fg:cartoon_outflow} shows an schematic drawing of this scenario. Assuming that the outflow is bi-conical with opening half-angle $\psi$ and hence is spread over the entirety of the regions illuminated by the accretion disk through the torus openings, a more conservative estimate on the column density can be achieved by using the luminosity of the [O{\sc iii}] outflow measured by \citet{finnerty20} of $\log L_{\rm{[OIII]}}/L_{\odot} = 11.5$. It can be shown that the mean electron density in this case would be given by
\begin{equation}\label{eq:mean_ne_bicone}
    {\langle}n_e{\rangle}\ \approx\ 13~{\rm cm^{-3}}~ f^{-1/2}~ (1-\cos\psi)^{-1/2}~ \left(\frac{L_{\rm{[OIII]}}}{10^{11.5}L_{\odot}}\right)^{1/2},
\end{equation}
\noindent where $f$ is the filling factor of the outflow. We have made the same assumptions of \citet{jun20} and \citet{finnerty20} that ${\langle}n_e{\rangle}^2/{\langle}n_e^2{\rangle}\approx 10^{[O/H]}$ (or, equivalently, that the metallicity is solar and that ${\langle}n_e^2{\rangle}\sim {\langle}n_e{\rangle}^2$), that the electron temperature is $\approx$10,000~K and that the gas has the same hydrogen and helium composition as before \citep[see also][]{carniani15}. We note that for that electron temperature, the [O{\sc iii}] emissivity is relatively insensitive to the electron density, showing a variation of less than 3\% between $n_e=1$ and $10,000~\rm cm^{-3}$ according to the software {\tt{PyNeb}} \citep{luridiana15}, so we have assumed the same value used by \citet{carniani15} of $3.4\times 10^{-21}~\rm erg~\rm s^{-1}~\rm cm^{3}$ for consistency. Following the same argument as in the previous paragraph to estimate a lower bound on the column density based on the mean electron density, we find that $N_{\rm H}\ge 1\times 10^{23}~\rm cm^{-2} (1-\cos\psi)^{-1/2}$. Using the median gas-to-dust ratio of \citet{maiolino01}, we find $E(B-V)\ge 1.5~\rm mag$ for the largest possible value of $\psi=\pi/2$, which would still imply an obscured view to the accretion disk through the outflow. We note that the electron densities implied by equation (\ref{eq:mean_ne_bicone}) for $\psi\gtrsim 5~\rm deg$ and $f\sim 1$ are well below the typical values of $\sim 500~\rm cm^{-3}$ found by, e.g., \citet{karouzos16} in AGN outflows, and well below the estimates of \citet{jun20} of ${\langle}n_e{\rangle}\sim 1100$ and $600~\rm cm^{-3}$ for outflows in two Hot DOGs using the [S{\sc ii}] doublet. This implies that either the opening angle is small, making the outflow typical dust reddening much larger than $1.5~\rm mag$ for lines of sight through the outflow, or that the outflow has a low filling factor, in which case obscuration might vary significantly for different lines of sight. A low filling factor might indeed be necessary to explain the high observed $L_{[OIII]}$, although we cannot disregard the possibility that the intrinsic $L_{[OIII]}$ could be larger due to obscuration. Combined with the uncertainties inherent to some of the assumptions made (as discussed in the previous paragraph), quantitative predictions are inaccurate. However, given the analysis presented here, we can qualitatively conclude that the outflow is likely optically thick at UV and optical wavelengths for most lines of sight.

\citet{zubko00} studied the polarization due to scattering by dust grains and showed that the degree of polarization from an optically thick dust sphere is almost unchanged from that of optically thin dust (see their Fig. 17). They also showed that at a rest-frame wavelength of 1600\AA\ the fraction of the scattered flux for an optically thick dust sphere rises with the scattering angle (see their Fig. 16), and is maximum at $\theta = 180~\rm deg$, when our line of sight looks directly at the illuminated side of the sphere (i.e., back-scattering). Considering this, we propose the following scenario. We assume the AGN in W0116--0505 is launching a bi-conical outflow. We also assume the bi-conical outflow has an $N_{\rm H}^{\rm out}$ consistent with the lower limits estimated above, and that the ISM outside of the outflow has a much lower column density. We propose then that the scattered flux we observe comes from back-scattered light on the surface, primarily at the base, of the receding outflow. Scattered light from the inner regions of the outflow is not dominant at 1600~\AA\ due to both dust reddening and multiple scattering events diluting the signal, although we note that the dust in the outflows cannot be too optically thick far from the base in order to permit us observing the blue-shifted [O{\sc iii}] emission from the approaching outflow that is very prominent in the spectrum of W0116--0505 \citep[see][]{finnerty20}.

This potential scenario has the interesting property that even for the lines of sight that go through the torus openings, W0116--0505 would likely be observed as a highly obscured quasar or, more specifically, a Hot DOG. In the next section we speculate how BHDs might be related to other populations of obscured hyper-luminous quasars in the literature. 

\subsection{A Potential Relation Between BHDs, ERQs and Heavily Reddenned Type 1 Quasars}

In \S\ref{ssec:outflow} we proposed that the blue excess in W0116--0505 might be caused by accretion disk light scattered by the massive ionized gas outflow known in the system. We also found that, in this scenario, an observer looking through the approaching outflow, and hence through the opening in the torus, would likely still observe an obscured quasar due to the optical thickness of the outflow. Specifically, we estimated that, in the best of conditions, a typical line of sight through the outflow would observe the central engine with a reddening of $E(B-V)\ge 1.5~\rm mag$ (although with considerable uncertainty). As the outflow expands, however, the column density will decrease, eventually allowing a significant number of lines of sight with a direct view to the accretion disk. Recently, \citet{banerji15} presented a population of hyper-luminous heavily reddened type 1 quasars at $z>2$ with $0.5 \lesssim E(B-V) \lesssim 1.5$. We speculate that, if the scenario in W0116--0505 of a bi-conical outflow is correct, W0116--0505 could potentially be a precursor to these heavily reddened type 1 quasars. Recently \citet{temple19} has shown that heavily reddened type 1 quasars also have fast ionized gas outflows. None of the objects they studied, however, had as high a velocity dispersion as the FWHM $4200\pm 100~\rm km~\rm s^{-1}$ found in W0116--0505 by \citet{finnerty20}, and all objects in \citet{temple19} had lower [O{\sc iii}] luminosities by 1 to 2 orders of magnitude compared to W0116--0505, suggesting the outflowing gas is more diffuse. \citet{lansbury20} finds the hydrogen column densities obscuring the X-rays in heavily reddened type 1 quasar are in the range of $1-8\times 10^{22}~\rm cm^{-2}$, considerably below the nearly Compton-thick obscuration of Hot DOGs and BHDs \citep{stern14, piconcelli15, assef16, assef20, vito18}. This might be consistent with W0116--0505 tracing an earlier evolutionary stage than heavily reddened type 1 quasars.

We further speculate that this source of the blue excess may extend to all or most BHDs. In particular, we note that \citet{finnerty20} found outflow properties similar to W0116--0505 for W0220+0137. As for W0116--0505, \citet{assef20} concluded that the most likely source for the UV emission is scattered emission from the central engine for W0220+0137, suggesting that polarimetric observations should also identify a significant degree of linear polarization. As mentioned earlier, both targets are also classified as ERQs, and a spectropolarimetric study of ERQs by \citet{alexandroff18} identified a significant degree of polarization in their UV light. Furthermore, \citet{zakamska16} identified significant ionized gas outflows in some ERQs, similar to those found by \citet{finnerty20} for W0116--0505 and W0220+0137. Hence, ERQs may be closely related to BHDs, and both may relate to heavily reddened type 1 hyper-luminous quasars. 

We suggest that Hot DOGs, ERQs and heavily reddened type 1 quasars are related to each other through both evolution and line of sight obscuration. Specifically, we hypothesize that Hot DOGs represent the earliest stage of these hyper-luminous objects, with the central engine completely enshrouded in dust. With no light escaping from the central engine into the ISM, their UV emission is dominated by starlight from the host galaxy \citep[see][]{wu12, assef15}, leading to relatively faint observed-frame optical fluxes. At some point in their evolution, these objects start driving massive, fast outflows, leading to the formation of large openings in their dust torus. In the early stages of this outflow phase, the objects are observed as BHDs/ERQs when observed along lines of sight that allow a view to the receding outflow and hence maximize the scattered flux. In the later stages of this outflow phase, once the outflow has expanded and hence lowered its column density, an observer looking through the outflow would see the object as a heavily reddened type 1 hyper-luminous quasar. The picture we have outlined here is qualitatively similar to that suggested by, e.g., \citet{hopkins08}, so we can also hypothesize that once the outflowing material is cleared out of the ISM, these objects will transition to regular type 1 quasars with no obscuration when observed through the openings in the dust torus. A similar scenario has been recently suggested by \citet{lansbury20} and \citet{jun21} based on estimates of the radiation feedback in luminous obscured quasars. Recently, \citet{diaz-santos21} has suggested that the availability of gas at large scales may lead to Hot DOGs being a recurring phase throughout the evolution of massive galaxies, with each episode perhaps leading to a BHD/ERQ/heavily reddened type 1 quasar phase as well, while the massive outflows slowly deplete the gas available in the ISM of their host galaxies.

Further polarimetric observations of more BHDs and ERQs will be needed to better constrain models for the light scattering and enable stronger connections to be drawn between these populations. {\it{JWST}}/NIRSPEC IFU observations may be able to determine the size and profile of the outflow in W0116--0505 as well as in other BHDs, ERQs, Hot DOGs and heavily reddened type 1 quasars. Additional constraints on the relationship between these populations may come from the studies of their environments, as those would remain stable through these relatively fast transitions.

\section{Conclusions}\label{sec:conclusions}

We have presented imaging polarimetry observations in the $R_{\rm Special}$ broad-band using the VLT/FORS2 instrument of W0116--0505 at $z=3.173$, identified by \citet{assef20} as a BHD. We measure a linear polarization of $p=10.8\pm 1.9\%$ with a polarization angle of $\chi = 74\pm 9~\rm deg$. This high degree of polarization confirms the scattered light scenario, and effectively discards an extreme, unobscured starburst as the primary source of the observed excess UV emission in W0116--0505. 

We discuss the implications for the dust obscuration geometry and the properties of the scattering medium in the context of a simple model in which the central engine is covered by a dust torus with two polar openings through which light can escape and illuminate an ISM that is optically thin for scattered light. For this model we find that both Thomson scattering and optically thin dust scattering can reproduce the observed linear polarization and total scattered flux for a wide range of ISM column densities and torus geometries. Optically thin dust scattering is much more likely than Thomson scattering because a dust-free scattering medium would only be expected within the dust sublimation radius of the AGN. 

However, this is not a unique scenario, and we also discuss the possibility that the scattering medium is a bi-conical dusty outflow aligned with the torus openings. In this scenario the scattered flux might be dominated by light back-scattered off the base of the outflow receding from us. In this scenario, once the outflow expands, observers with a direct line of sight to the accretion disk through the dusty outflow might see a reddened AGN, possibly consistent with the heavily reddened type 1 hyper-luminous quasars identified by \citet{banerji15}. We hypothesize that BHDs are closely related to ERQs, and may correspond to the same objects identified by \citet{banerji15} at a slightly earlier evolutionary stage. We speculate that there could be an evolutionary sequence in which BHDs/ERQs/heavily reddened type 1 hyper-luminous quasars are the intermediate stage between Hot DOGs without blue-excess and traditional type 1 quasars of similar luminosities.

\begin{acknowledgments}
We thank Aleksandar Cikota for helping us with the reduction of the FORS2 imaging polarimetry data and Matthew Temple and Aaron Barth for comments and suggestions that helped improve our work. We also thank the anonymous referee for all of their comments and suggestions on the submitted manuscript. RJA was supported by FONDECYT grant number 1191124 and by ANID BASAL project FB210003. FEB acknowledges support from ANID-Chile BASAL AFB-170002, ACE210002, and FB210003, FONDECYT Regular 1200495 and 1190818, and Millennium Science Initiative Program  – ICN12\_009. HDJ was supported by the National Research Foundation of Korea (NRF) grant 2022R1C1C2013543 funded by the Ministry of Science and ICT (MSIT) of Korea. CWT acknowledges support from the NSFC grant 11973051. Based on observations collected at the European Organisation for Astronomical Research in the Southern Hemisphere under ESO programme 106.218J.001. This research uses data products from the Wide-field Infrared Survey Explorer, which is a joint project of the University of California, Los Angeles, and the Jet Propulsion Laboratory (JPL)/California Institute of Technology (Caltech), funded by the National Aeronautics and Space Administration (NASA). Portions of this research were carried out at JPL/Caltech under a contract with NASA. This research made use of Photutils, an Astropy package for detection and photometry of astronomical sources \citep{bradley19}. This research is based on observations made with the NASA/ESA Hubble Space Telescope obtained from the Space Telescope Science Institute, which is operated by the Association of Universities for Research in Astronomy, Inc., under NASA contract NAS 5–26555. These observations are associated with program 14358. Funding for SDSS-III has been provided by the Alfred P. Sloan Foundation, the Participating Institutions, the National Science Foundation, and the U.S. Department of Energy Office of Science. 
\end{acknowledgments}


\begin{thebibliography}{}
    \expandafter\ifx\csname natexlab\endcsname\relax\def\natexlab#1{#1}\fi
    \providecommand{\url}[1]{\href{#1}{#1}}
    \providecommand{\dodoi}[1]{doi:~\href{http://doi.org/#1}{\nolinkurl{#1}}}
    \providecommand{\doeprint}[1]{\href{http://ascl.net/#1}{\nolinkurl{http://ascl.net/#1}}}
    \providecommand{\doarXiv}[1]{\href{https://arxiv.org/abs/#1}{\nolinkurl{https://arxiv.org/abs/#1}}}
    
    \bibitem[{{Alexandroff} {et~al.}(2018){Alexandroff}, {Zakamska}, {Barth},
      {Hamann}, {Strauss}, {Krolik}, {Greene}, {P{\^a}ris}, \&
      {Ross}}]{alexandroff18}
    {Alexandroff}, R.~M., {Zakamska}, N.~L., {Barth}, A.~J., {et~al.} 2018, \mnras,
      479, 4936, \dodoi{10.1093/mnras/sty1685}
    
    \bibitem[{{Alonso-Herrero} {et~al.}(2021){Alonso-Herrero},
      {Garc{\'\i}a-Burillo}, {H{\"o}nig}, {Garc{\'\i}a-Bernete}, {Ramos Almeida},
      {Gonz{\'a}lez-Mart{\'\i}n}, {L{\'o}pez-Rodr{\'\i}guez}, {Boorman}, {Bunker},
      {Burtscher}, {Combes}, {Davies}, {D{\'\i}az-Santos}, {Gandhi},
      {Garc{\'\i}a-Lorenzo}, {Hicks}, {Hunt}, {Ichikawa}, {Imanishi}, {Izumi},
      {Labiano}, {Levenson}, {Packham}, {Pereira-Santaella}, {Ricci}, {Rigopoulou},
      {Roche}, {Rosario}, {Rouan}, {Shimizu}, {Stalevski}, {Wada}, \&
      {Williamson}}]{alonsoherrero21}
    {Alonso-Herrero}, A., {Garc{\'\i}a-Burillo}, S., {H{\"o}nig}, S.~F., {et~al.}
      2021, \aap, 652, A99, \dodoi{10.1051/0004-6361/202141219}
    
    \bibitem[{{Antonucci}(1993)}]{antonucci93}
    {Antonucci}, R. 1993, \araa, 31, 473,
      \dodoi{10.1146/annurev.aa.31.090193.002353}
    
    \bibitem[{{Antonucci} \& {Miller}(1985)}]{antonucci85}
    {Antonucci}, R.~R.~J., \& {Miller}, J.~S. 1985, \apj, 297, 621,
      \dodoi{10.1086/163559}
    
    \bibitem[{{Assef} {et~al.}(2010){Assef}, {Kochanek}, {Brodwin}, {Cool},
      {Forman}, {Gonzalez}, {Hickox}, {Jones}, {Le Floc'h}, {Moustakas}, {Murray},
      \& {Stern}}]{assef10}
    {Assef}, R.~J., {Kochanek}, C.~S., {Brodwin}, M., {et~al.} 2010, \apj, 713,
      970, \dodoi{10.1088/0004-637X/713/2/970}
    
    \bibitem[{{Assef} {et~al.}(2013){Assef}, {Stern}, {Kochanek}, {Blain},
      {Brodwin}, {Brown}, {Donoso}, {Eisenhardt}, {Jannuzi}, {Jarrett}, {Stanford},
      {Tsai}, {Wu}, \& {Yan}}]{assef13}
    {Assef}, R.~J., {Stern}, D., {Kochanek}, C.~S., {et~al.} 2013, \apj, 772, 26,
      \dodoi{10.1088/0004-637X/772/1/26}
    
    \bibitem[{{Assef} {et~al.}(2015){Assef}, {Eisenhardt}, {Stern}, {Tsai}, {Wu},
      {Wylezalek}, {Blain}, {Bridge}, {Donoso}, {Gonzales}, {Griffith}, \&
      {Jarrett}}]{assef15}
    {Assef}, R.~J., {Eisenhardt}, P.~R.~M., {Stern}, D., {et~al.} 2015, \apj, 804,
      27, \dodoi{10.1088/0004-637X/804/1/27}
    
    \bibitem[{{Assef} {et~al.}(2016){Assef}, {Walton}, {Brightman}, {Stern},
      {Alexander}, {Bauer}, {Blain}, {Diaz-Santos}, {Eisenhardt}, {Finkelstein},
      {Hickox}, {Tsai}, \& {Wu}}]{assef16}
    {Assef}, R.~J., {Walton}, D.~J., {Brightman}, M., {et~al.} 2016, \apj, 819,
      111, \dodoi{10.3847/0004-637X/819/2/111}
    
    \bibitem[{{Assef} {et~al.}(2020){Assef}, {Brightman}, {Walton}, {Stern},
      {Bauer}, {Blain}, {D{\'\i}az-Santos}, {Eisenhardt}, {Hickox}, {Jun},
      {Psychogyios}, {Tsai}, \& {Wu}}]{assef20}
    {Assef}, R.~J., {Brightman}, M., {Walton}, D.~J., {et~al.} 2020, \apj, 897,
      112, \dodoi{10.3847/1538-4357/ab9814}
    
    \bibitem[{{Banerji} {et~al.}(2015){Banerji}, {Alaghband-Zadeh}, {Hewett}, \&
      {McMahon}}]{banerji15}
    {Banerji}, M., {Alaghband-Zadeh}, S., {Hewett}, P.~C., \& {McMahon}, R.~G.
      2015, \mnras, 447, 3368, \dodoi{10.1093/mnras/stu2649}
    
    \bibitem[{{Berriman} {et~al.}(1990){Berriman}, {Schmidt}, {West}, \&
      {Stockman}}]{berriman90}
    {Berriman}, G., {Schmidt}, G.~D., {West}, S.~C., \& {Stockman}, H.~S. 1990,
      \apjs, 74, 869, \dodoi{10.1086/191523}
    
    \bibitem[{{Bradley} {et~al.}(2019){Bradley}, {Sipocz}, {Robitaille},
      {Tollerud}, {Vin{\'\i}cius}, {Deil}, {Barbary}, {Busko}, {G{\"u}nther},
      {Cara}, {Wilson}, {Conseil}, {Droettboom}, {Bostroem}, {Bray}, {Andersen
      Bratholm}, {Craig}, {Barentsen}, {Pascual}, {Lim}, {Donath}, {Greco},
      {Perren}, {Kerzendorf}, {De Val-Borro}, {Dencheva}, {De Albernaz Ferreira},
      {Souchereau}, {D'Eugenio}, \& {Weaver}}]{bradley19}
    {Bradley}, L., {Sipocz}, B., {Robitaille}, T., {et~al.} 2019,
      {astropy/photutils: v0.7}, v0.7,  Zenodo, \dodoi{10.5281/zenodo.3368647}
    
    \bibitem[{{Cardelli} {et~al.}(1989){Cardelli}, {Clayton}, \&
      {Mathis}}]{cardelli89}
    {Cardelli}, J.~A., {Clayton}, G.~C., \& {Mathis}, J.~S. 1989, \apj, 345, 245,
      \dodoi{10.1086/167900}
    
    \bibitem[{{Carniani} {et~al.}(2015){Carniani}, {Marconi}, {Maiolino},
      {Balmaverde}, {Brusa}, {Cano-D{\'\i}az}, {Cicone}, {Comastri}, {Cresci},
      {Fiore}, {Feruglio}, {La Franca}, {Mainieri}, {Mannucci}, {Nagao}, {Netzer},
      {Piconcelli}, {Risaliti}, {Schneider}, \& {Shemmer}}]{carniani15}
    {Carniani}, S., {Marconi}, A., {Maiolino}, R., {et~al.} 2015, \aap, 580, A102,
      \dodoi{10.1051/0004-6361/201526557}
    
    \bibitem[{{Coleman} {et~al.}(1980){Coleman}, {Wu}, \& {Weedman}}]{cww80}
    {Coleman}, G.~D., {Wu}, C.~C., \& {Weedman}, D.~W. 1980, \apjs, 43, 393,
      \dodoi{10.1086/190674}
    
    \bibitem[{{D{\'\i}az-Santos} {et~al.}(2016){D{\'\i}az-Santos}, {Assef},
      {Blain}, {Tsai}, {Aravena}, {Eisenhardt}, {Wu}, {Stern}, \&
      {Bridge}}]{diaz-santos16}
    {D{\'\i}az-Santos}, T., {Assef}, R.~J., {Blain}, A.~W., {et~al.} 2016, \apjl,
      816, L6, \dodoi{10.3847/2041-8205/816/1/L6}
    
    \bibitem[{{D{\'\i}az-Santos} {et~al.}(2018){D{\'\i}az-Santos}, {Assef},
      {Blain}, {Aravena}, {Stern}, {Tsai}, {Eisenhardt}, {Wu}, {Jun}, {Dibert},
      {Inami}, {Lansbury}, \& {Leclercq}}]{diaz-santos18}
    ---. 2018, Science, 362, 1034, \dodoi{10.1126/science.aap7605}
    
    \bibitem[{{Diaz-Santos} {et~al.}(2021){Diaz-Santos}, {Assef}, {Eisenhardt},
      {Jun}, {Jones}, {Blain}, {Stern}, {Aravena}, {Tsai}, {Lake}, {Wu}, \&
      {Gonzalez-Lopez}}]{diaz-santos21}
    {Diaz-Santos}, T., {Assef}, R.~J., {Eisenhardt}, P. R.~M., {et~al.} 2021, arXiv
      e-prints, arXiv:2104.09495.
    \newblock \doarXiv{2104.09495}
    
    \bibitem[{{Draine}(2003)}]{draine03}
    {Draine}, B.~T. 2003, \apj, 598, 1017, \dodoi{10.1086/379118}
    
    \bibitem[{{Eisenhardt} {et~al.}(in prep.)}]{eisenhardt22}
    {Eisenhardt}, P. R.~M., {et~al.} in prep.
    
    \bibitem[{{Eisenhardt} {et~al.}(2012){Eisenhardt}, {Wu}, {Tsai}, {Assef},
      {Benford}, {Blain}, {Bridge}, {Condon}, {Cushing}, {Cutri}, {Evans},
      {Gelino}, {Griffith}, {Grillmair}, {Jarrett}, {Lonsdale}, {Masci}, {Mason},
      {Petty}, {Sayers}, {Stanford}, {Stern}, {Wright}, \& {Yan}}]{eisenhardt12}
    {Eisenhardt}, P. R.~M., {Wu}, J., {Tsai}, C.-W., {et~al.} 2012, \apj, 755, 173,
      \dodoi{10.1088/0004-637X/755/2/173}
    
    \bibitem[{{Eisenstein} {et~al.}(2011){Eisenstein}, {Weinberg}, {Agol},
      {Aihara}, {Allende Prieto}, {Anderson}, {Arns}, {Aubourg}, {Bailey},
      {Balbinot}, \& et~al.}]{eisenstein11}
    {Eisenstein}, D.~J., {Weinberg}, D.~H., {Agol}, E., {et~al.} 2011, \aj, 142,
      72, \dodoi{10.1088/0004-6256/142/3/72}
    
    \bibitem[{{Fan} {et~al.}(2018){Fan}, {Knudsen}, {Fogasy}, \& {Drouart}}]{fan18}
    {Fan}, L., {Knudsen}, K.~K., {Fogasy}, J., \& {Drouart}, G. 2018, \apjl, 856,
      L5, \dodoi{10.3847/2041-8213/aab496}
    
    \bibitem[{{Fan} {et~al.}(2016){Fan}, {Han}, {Fang}, {Gao}, {Zhang}, {Jiang},
      {Wu}, {Yang}, \& {Li}}]{fan16}
    {Fan}, L., {Han}, Y., {Fang}, G., {et~al.} 2016, \apjl, 822, L32,
      \dodoi{10.3847/2041-8205/822/2/L32}
    
    \bibitem[{{Farrah} {et~al.}(2017){Farrah}, {Petty}, {Connolly}, {Blain},
      {Efstathiou}, {Lacy}, {Stern}, {Lake}, {Jarrett}, {Bridge}, {Eisenhardt},
      {Benford}, {Jones}, {Tsai}, {Assef}, {Wu}, \& {Moustakas}}]{farrah17}
    {Farrah}, D., {Petty}, S., {Connolly}, B., {et~al.} 2017, \apj, 844, 106,
      \dodoi{10.3847/1538-4357/aa78f2}
    
    \bibitem[{{Finnerty} {et~al.}(2020){Finnerty}, {Larson}, {Soifer}, {Armus},
      {Matthews}, {Jun}, {Moon}, {Melbourne}, {Gomez}, {Tsai}, {D{\'\i}az-Santos},
      {Eisenhardt}, \& {Cushing}}]{finnerty20}
    {Finnerty}, L., {Larson}, K., {Soifer}, B.~T., {et~al.} 2020, \apj, 905, 16,
      \dodoi{10.3847/1538-4357/abc3bf}
    
    \bibitem[{{Fossati} {et~al.}(2007){Fossati}, {Bagnulo}, {Mason}, \& {Landi
      Degl'Innocenti}}]{fossati07}
    {Fossati}, L., {Bagnulo}, S., {Mason}, E., \& {Landi Degl'Innocenti}, E. 2007,
      in Astronomical Society of the Pacific Conference Series, Vol. 364, The
      Future of Photometric, Spectrophotometric and Polarimetric Standardization,
      ed. C.~{Sterken}, 503
    
    \bibitem[{{Gonz{\'a}lez-Gait{\'a}n} {et~al.}(2020){Gonz{\'a}lez-Gait{\'a}n},
      {Mour{\~a}o}, {Patat}, {Anderson}, {Cikota}, {Wiersema}, {Higgins}, \&
      {Silva}}]{gonzalez20}
    {Gonz{\'a}lez-Gait{\'a}n}, S., {Mour{\~a}o}, A.~M., {Patat}, F., {et~al.} 2020,
      \aap, 634, A70, \dodoi{10.1051/0004-6361/201936379}
    
    \bibitem[{{Gordon} \& {Clayton}(1998)}]{gordon98}
    {Gordon}, K.~D., \& {Clayton}, G.~C. 1998, \apj, 500, 816,
      \dodoi{10.1086/305774}
    
    \bibitem[{{Goulding} {et~al.}(2018){Goulding}, {Zakamska}, {Alexandroff},
      {Assef}, {Banerji}, {Hamann}, {Wylezalek}, {Brandt}, {Greene}, {Lansbury},
      {P{\^a}ris}, {Richards}, {Stern}, \& {Strauss}}]{goulding18}
    {Goulding}, A.~D., {Zakamska}, N.~L., {Alexandroff}, R.~M., {et~al.} 2018,
      \apj, 856, 4, \dodoi{10.3847/1538-4357/aab040}
    
    \bibitem[{{G{\"u}ver} \& {{\"O}zel}(2009)}]{guver09}
    {G{\"u}ver}, T., \& {{\"O}zel}, F. 2009, \mnras, 400, 2050,
      \dodoi{10.1111/j.1365-2966.2009.15598.x}
    
    \bibitem[{{Hamann} {et~al.}(2017){Hamann}, {Zakamska}, {Ross}, {Paris},
      {Alexandroff}, {Villforth}, {Richards}, {Herbst}, {Brandt}, {Cook}, {Denney},
      {Greene}, {Schneider}, \& {Strauss}}]{hamann17}
    {Hamann}, F., {Zakamska}, N.~L., {Ross}, N., {et~al.} 2017, \mnras, 464, 3431,
      \dodoi{10.1093/mnras/stw2387}
    
    \bibitem[Heiles(2000)]{heiles00} 
        Heiles, C.\ 2000, \aj, 119, 923. doi:10.1086/301236
    
    \bibitem[{{Hines} {et~al.}(1995){Hines}, {Schmidt}, {Smith}, {Cutri}, \&
      {Low}}]{hines95}
    {Hines}, D.~C., {Schmidt}, G.~D., {Smith}, P.~S., {Cutri}, R.~M., \& {Low},
      F.~J. 1995, \apjl, 450, L1, \dodoi{10.1086/309658}
    
    \bibitem[{{Hopkins} {et~al.}(2008){Hopkins}, {Hernquist}, {Cox}, \&
      {Kere{\v{s}}}}]{hopkins08}
    {Hopkins}, P.~F., {Hernquist}, L., {Cox}, T.~J., \& {Kere{\v{s}}}, D. 2008,
      \apjs, 175, 356, \dodoi{10.1086/524362}
    
    \bibitem[{{Jones} {et~al.}(2014){Jones}, {Blain}, {Stern}, {Assef}, {Bridge},
      {Eisenhardt}, {Petty}, {Wu}, {Tsai}, {Cutri}, {Wright}, \& {Yan}}]{jones14}
    {Jones}, S.~F., {Blain}, A.~W., {Stern}, D., {et~al.} 2014, \mnras, 443, 146,
      \dodoi{10.1093/mnras/stu1157}
    
    \bibitem[{{Jun} {et~al.}(2021){Jun}, {Assef}, {Carroll}, {Hickox}, {Kim},
      {Lee}, {Ricci}, \& {Stern}}]{jun21}
    {Jun}, H.~D., {Assef}, R.~J., {Carroll}, C.~M., {et~al.} 2021, \apj, 906, 21,
      \dodoi{10.3847/1538-4357/abc629}
    
    \bibitem[{{Jun} {et~al.}(2020){Jun}, {Assef}, {Bauer}, {Blain},
      {D{\'\i}az-Santos}, {Eisenhardt}, {Stern}, {Tsai}, {Wright}, \& {Wu}}]{jun20}
    {Jun}, H.~D., {Assef}, R.~J., {Bauer}, F.~E., {et~al.} 2020, \apj, 888, 110,
      \dodoi{10.3847/1538-4357/ab5e7b}
    
    \bibitem[{{Karouzos} {et~al.}(2016){Karouzos}, {Woo}, \& {Bae}}]{karouzos16}
    {Karouzos}, M., {Woo}, J.-H., \& {Bae}, H.-J. 2016, \apj, 833, 171,
      \dodoi{10.3847/1538-4357/833/2/171}
    
    \bibitem[Lansbury et al.(2020)]{lansbury20} 
        Lansbury, G.~B., Banerji, M., Fabian, A.~C., et al.\ 2020, \mnras, 495, 2652. doi:10.1093/mnras/staa1220
    
    \bibitem[{{Luridiana} {et~al.}(2015){Luridiana}, {Morisset}, \&
      {Shaw}}]{luridiana15}
    {Luridiana}, V., {Morisset}, C., \& {Shaw}, R.~A. 2015, \aap, 573, A42,
      \dodoi{10.1051/0004-6361/201323152}
    
    \bibitem[MacLeod et al.(2010)]{macleod10} 
    MacLeod, C.~L., Ivezi{\'c}, {\v{Z}}., Kochanek, C.~S., et al.\ 2010, \apj, 
        721, 1014, doi:10.1088/0004-637X/721/2/1014
    
    \bibitem[{{Maiolino} {et~al.}(2001){Maiolino}, {Marconi}, {Salvati},
      {Risaliti}, {Severgnini}, {Oliva}, {La Franca}, \& {Vanzi}}]{maiolino01}
    {Maiolino}, R., {Marconi}, A., {Salvati}, M., {et~al.} 2001, \aap, 365, 28,
      \dodoi{10.1051/0004-6361:20000177}
    
    \bibitem[{{Mathis} {et~al.}(1977){Mathis}, {Rumpl}, \& {Nordsieck}}]{mathis77}
    {Mathis}, J.~S., {Rumpl}, W., \& {Nordsieck}, K.~H. 1977, \apj, 217, 425,
      \dodoi{10.1086/155591}
    
    \bibitem[{{McCully} {et~al.}(2018){McCully}, {Crawford}, {Kovacs}, {Tollerud},
      {Betts}, {Bradley}, {Craig}, {Turner}, {Streicher}, {Sipocz}, {Robitaille},
      \& {Deil}}]{mccully18}
    {McCully}, C., {Crawford}, S., {Kovacs}, G., {et~al.} 2018,
      {Astropy/Astroscrappy: V1.0.5 Zenodo Release}, v1.0.5,  Zenodo,
      \dodoi{10.5281/zenodo.1482019}
    
    \bibitem[{{Miller} {et~al.}(1991){Miller}, {Goodrich}, \& {Mathews}}]{miller91}
    {Miller}, J.~S., {Goodrich}, R.~W., \& {Mathews}, W.~G. 1991, \apj, 378, 47,
      \dodoi{10.1086/170406}
    
    \bibitem[{{Nenkova} {et~al.}(2008){Nenkova}, {Sirocky}, {Nikutta},
      {Ivezi{\'c}}, \& {Elitzur}}]{nenkova08}
    {Nenkova}, M., {Sirocky}, M.~M., {Nikutta}, R., {Ivezi{\'c}}, {\v{Z}}., \&
      {Elitzur}, M. 2008, \apj, 685, 160, \dodoi{10.1086/590483}
    
    \bibitem[{{Piconcelli} {et~al.}(2015){Piconcelli}, {Vignali}, {Bianchi},
      {Zappacosta}, {Fritz}, {Lanzuisi}, {Miniutti}, {Bongiorno}, {Feruglio},
      {Fiore}, \& {Maiolino}}]{piconcelli15}
    {Piconcelli}, E., {Vignali}, C., {Bianchi}, S., {et~al.} 2015, \aap, 574, L9,
      \dodoi{10.1051/0004-6361/201425324}
    
    \bibitem[{{Planck Collaboration} {et~al.}(2016){Planck Collaboration}, {Ade},
      {Aghanim}, {Arnaud}, {Ashdown}, {Aumont}, {Baccigalupi}, {Banday},
      {Barreiro}, {Bartlett}, \& et~al.}]{planck16}
    {Planck Collaboration}, {Ade}, P.~A.~R., {Aghanim}, N., {et~al.} 2016, \aap,
      594, A13, \dodoi{10.1051/0004-6361/201525830}
    
    \bibitem[Rakshit et al.(2020)]{rakshit20} 
    Rakshit, S., Stalin, C.~S., \& Kotilainen, J.\ 2020, \apjs, 249, 17, 
        doi:10.3847/1538-4365/ab99c5
    
    \bibitem[{{Roseboom} {et~al.}(2013){Roseboom}, {Lawrence}, {Elvis}, {Petty},
      {Shen}, \& {Hao}}]{roseboom13}
    {Roseboom}, I.~G., {Lawrence}, A., {Elvis}, M., {et~al.} 2013, \mnras, 429,
      1494, \dodoi{10.1093/mnras/sts441}
    
    \bibitem[{{Rupke} {et~al.}(2017){Rupke}, {G{\"u}ltekin}, \&
      {Veilleux}}]{rupke17}
    {Rupke}, D. S.~N., {G{\"u}ltekin}, K., \& {Veilleux}, S. 2017, \apj, 850, 40,
      \dodoi{10.3847/1538-4357/aa94d1}
    
    \bibitem[{{Schindler} {et~al.}(2019){Schindler}, {Fan}, {McGreer}, {Yang},
      {Wang}, {Green}, {Fynbo}, {Krogager}, {Green}, {Huang}, {Kadowaki}, {Patej},
      {Wu}, \& {Yue}}]{schindler19}
    {Schindler}, J.-T., {Fan}, X., {McGreer}, I.~D., {et~al.} 2019, \apj, 871, 258,
      \dodoi{10.3847/1538-4357/aaf86c}
    
    \bibitem[{{Schlesinger} {et~al.}(2009){Schlesinger}, {Pogge}, {Martini},
      {Shields}, \& {Fields}}]{schlesinger09}
    {Schlesinger}, K., {Pogge}, R.~W., {Martini}, P., {Shields}, J.~C., \&
      {Fields}, D. 2009, \apj, 699, 857, \dodoi{10.1088/0004-637X/699/1/857}
    
    \bibitem[{{Schmidt} \& {Green}(1983)}]{schmidt83}
    {Schmidt}, M., \& {Green}, R.~F. 1983, \apj, 269, 352, \dodoi{10.1086/161048}
    
    \bibitem[{{Smith} {et~al.}(2000){Smith}, {Schmidt}, {Hines}, {Cutri}, \&
      {Nelson}}]{smith00}
    {Smith}, P.~S., {Schmidt}, G.~D., {Hines}, D.~C., {Cutri}, R.~M., \& {Nelson},
      B.~O. 2000, \apjl, 545, L19, \dodoi{10.1086/317329}
    
    \bibitem[{{Stalevski} {et~al.}(2017){Stalevski}, {Asmus}, \&
      {Tristram}}]{stalevski17}
    {Stalevski}, M., {Asmus}, D., \& {Tristram}, K. R.~W. 2017, \mnras, 472, 3854,
      \dodoi{10.1093/mnras/stx2227}
    
    \bibitem[{{Stern} {et~al.}(2014){Stern}, {Lansbury}, {Assef}, {Brandt},
      {Alexander}, {Ballantyne}, {Balokovi{\'c}}, {Bauer}, {Benford}, {Blain},
      {Boggs}, {Bridge}, {Brightman}, {Christensen}, {Comastri}, {Craig}, {Del
      Moro}, {Eisenhardt}, {Gandhi}, {Griffith}, {Hailey}, {Harrison}, {Hickox},
      {Jarrett}, {Koss}, {Lake}, {LaMassa}, {Luo}, {Tsai}, {Urry}, {Walton},
      {Wright}, {Wu}, {Yan}, \& {Zhang}}]{stern14}
    {Stern}, D., {Lansbury}, G.~B., {Assef}, R.~J., {et~al.} 2014, \apj, 794, 102,
      \dodoi{10.1088/0004-637X/794/2/102}
    
    \bibitem[{{Temple} {et~al.}(2019){Temple}, {Banerji}, {Hewett}, {Coatman},
      {Maddox}, \& {Peroux}}]{temple19}
    {Temple}, M.~J., {Banerji}, M., {Hewett}, P.~C., {et~al.} 2019, \mnras, 487,
      2594, \dodoi{10.1093/mnras/stz1420}
    
    \bibitem[Tonry et al.(2012)]{tonry12} 
    Tonry, J.~L., Stubbs, C.~W., Lykke, K.~R., et al.\ 2012, \apj, 750, 99,
        doi:10.1088/0004-637X/750/2/99
    
    \bibitem[{{Tsai} {et~al.}(2015){Tsai}, {Eisenhardt}, {Wu}, {Stern}, {Assef},
      {Blain}, {Bridge}, {Benford}, {Cutri}, {Griffith}, {Jarrett}, {Lonsdale},
      {Masci}, {Moustakas}, {Petty}, {Sayers}, {Stanford}, {Wright}, {Yan},
      {Leisawitz}, {Liu}, {Mainzer}, {McLean}, {Padgett}, {Skrutskie}, {Gelino},
      {Beichman}, \& {Juneau}}]{tsai15}
    {Tsai}, C.-W., {Eisenhardt}, P. R.~M., {Wu}, J., {et~al.} 2015, \apj, 805, 90,
      \dodoi{10.1088/0004-637X/805/2/90}
    
    \bibitem[Vanden Berk et al.(2004)]{vandenberk04} 
    Vanden Berk, D.~E., Wilhite, B.~C., Kron, R.~G., et al.\ 2004, \apj, 601, 692, 
    doi:10.1086/380563
    
    \bibitem[Vanden Berk et al.(2001)]{vandenberk01} 
    Vanden Berk, D.~E., Richards, G.~T., Bauer, A., et al.\ 2001, \aj, 122, 549, 
    doi:10.1086/321167
    
    \bibitem[{{van Dokkum}(2001)}]{vandokkum01}
    {van Dokkum}, P.~G. 2001, \pasp, 113, 1420, \dodoi{10.1086/323894}
    
    \bibitem[{{Vernet} {et~al.}(2001){Vernet}, {Fosbury}, {Villar-Mart{\'\i}n},
      {Cohen}, {Cimatti}, {di Serego Alighieri}, \& {Goodrich}}]{vernet01}
    {Vernet}, J., {Fosbury}, R.~A.~E., {Villar-Mart{\'\i}n}, M., {et~al.} 2001,
      \aap, 366, 7, \dodoi{10.1051/0004-6361:20000076}
    
    \bibitem[{{Vito} {et~al.}(2018){Vito}, {Brandt}, {Stern}, {Assef}, {Chen},
      {Brightman}, {Comastri}, {Eisenhardt}, {Garmire}, {Hickox}, {Lansbury},
      {Tsai}, {Walton}, \& {Wu}}]{vito18}
    {Vito}, F., {Brandt}, W.~N., {Stern}, D., {et~al.} 2018, \mnras, 474, 4528,
      \dodoi{10.1093/mnras/stx3120}
    
    \bibitem[{{Weingartner} \& {Draine}(2001)}]{weingartner01}
    {Weingartner}, J.~C., \& {Draine}, B.~T. 2001, \apj, 548, 296,
      \dodoi{10.1086/318651}
    
    \bibitem[{{Wright} {et~al.}(2010){Wright}, {Eisenhardt}, {Mainzer}, {Ressler},
      {Cutri}, {Jarrett}, {Kirkpatrick}, {Padgett}, {McMillan}, {Skrutskie},
      {Stanford}, {Cohen}, {Walker}, {Mather}, {Leisawitz}, {Gautier}, {McLean},
      {Benford}, {Lonsdale}, {Blain}, {Mendez}, {Irace}, {Duval}, {Liu}, {Royer},
      {Heinrichsen}, {Howard}, {Shannon}, {Kendall}, {Walsh}, {Larsen}, {Cardon},
      {Schick}, {Schwalm}, {Abid}, {Fabinsky}, {Naes}, \& {Tsai}}]{wright10}
    {Wright}, E.~L., {Eisenhardt}, P. R.~M., {Mainzer}, A.~K., {et~al.} 2010, \aj,
      140, 1868, \dodoi{10.1088/0004-6256/140/6/1868}
    
    \bibitem[{{Wu} {et~al.}(2012){Wu}, {Tsai}, {Sayers}, {Benford}, {Bridge},
      {Blain}, {Eisenhardt}, {Stern}, {Petty}, {Assef}, {Bussmann}, {Comerford},
      {Cutri}, {Evans}, {Griffith}, {Jarrett}, {Lake}, {Lonsdale}, {Rho},
      {Stanford}, {Weiner}, {Wright}, \& {Yan}}]{wu12}
    {Wu}, J., {Tsai}, C.-W., {Sayers}, J., {et~al.} 2012, \apj, 756, 96,
      \dodoi{10.1088/0004-637X/756/1/96}
    
    \bibitem[{{Wu} {et~al.}(2014){Wu}, {Bussmann}, {Tsai}, {Petric}, {Blain},
      {Eisenhardt}, {Bridge}, {Benford}, {Stern}, {Assef}, {Gelino}, {Moustakas},
      \& {Wright}}]{wu14}
    {Wu}, J., {Bussmann}, R.~S., {Tsai}, C.-W., {et~al.} 2014, \apj, 793, 8,
      \dodoi{10.1088/0004-637X/793/1/8}
    
    \bibitem[{{Zakamska} {et~al.}(2005){Zakamska}, {Schmidt}, {Smith}, {Strauss},
      {Krolik}, {Hall}, {Richards}, {Schneider}, {Brinkmann}, \&
      {Szokoly}}]{zakamska05}
    {Zakamska}, N.~L., {Schmidt}, G.~D., {Smith}, P.~S., {et~al.} 2005, \aj, 129,
      1212, \dodoi{10.1086/427543}
    
    \bibitem[{{Zakamska} {et~al.}(2016){Zakamska}, {Hamann}, {P{\^a}ris}, {Brandt},
      {Greene}, {Strauss}, {Villforth}, {Wylezalek}, {Alexandroff}, \&
      {Ross}}]{zakamska16}
    {Zakamska}, N.~L., {Hamann}, F., {P{\^a}ris}, I., {et~al.} 2016, \mnras, 459,
      3144, \dodoi{10.1093/mnras/stw718}
    
    \bibitem[{{Zubko} \& {Laor}(2000)}]{zubko00}
    {Zubko}, V.~G., \& {Laor}, A. 2000, \apjs, 128, 245, \dodoi{10.1086/313373}
    
\end{thebibliography}
\end{document}